%% file: main.tex
\pdfoutput=1
\documentclass{acmart}
\usepackage{graphicx}
\usepackage{multirow}

\usepackage{tikz}
\usetikzlibrary{arrows, arrows.spaced,calc, positioning,decorations.pathreplacing,angles,quotes,patterns}
\usepackage{pgfplots}
\pgfplotsset{compat=1.14}
\pgfplotsset{
    /pgfplots/ybar legend/.style={
    /pgfplots/legend image code/.code={%
       \draw[##1,/tikz/.cd,yshift=-0.25em]
        (0cm,0cm) rectangle (3pt,0.8em);},
   },
}
\usepgfplotslibrary{fillbetween, statistics}
\usepackage{tkz-euclide}

\usepackage[binary-units]{siunitx}
\usepackage[nointegrals]{wasysym}
\usepackage{varwidth}
\usepackage{mathtools}
\usepackage{amsmath,amsfonts}
\usepackage{todonotes}
\usepackage{subcaption}
\usepackage{placeins}
\usepackage{tabularx}
\newcolumntype{Y}{>{\centering\arraybackslash}X}

\usepackage{xspace}

\usepackage[noend]{algorithmic}
\usepackage{algorithm}

\usepackage{listings}
\newcommand{\deltat}{\ensuremath{\Delta\tau}}
\usepackage{mathtools}

\NewDocumentCommand{\R}{}{{\mathbb{R}}}

\NewDocumentCommand{\GFiring}{om}{{s_{#2}\IfValueT{#1}{^{#1}}}}
\NewDocumentCommand{\RVBound}{om}{{S_{#2}\IfValueT{#1}{^{#1}}}}

\NewDocumentCommand{\STDRegion}{}{{\symcal{R}}}

\NewDocumentCommand{\EntryMarking}{O{\STDRegion} m}{\operatorname{x^{#1}_{0, #2}}}
\NewDocumentCommand{\EntryTime}{O{\STDRegion}}{\operatorname{t^{#1}_0}}
\NewDocumentCommand{\ContinuousMarking}{o m}{\operatorname{x\IfValueT{#1}{^{#1}}_{#2}}}
\NewDocumentCommand{\Drift}{o m}{\operatorname{d\IfValueT{#1}{^{#1}}_{#2}}}
\NewDocumentCommand{\DiscreteMarking}{o m}{\operatorname{m\IfValueT{#1}{^{#1}}_{#2}}}

\DeclareMathOperator{\conv}{conv}

\NewDocumentCommand{\sat}{o}{\operatorname{Sat\IfValueT{#1}{^{#1}}}}
\NewDocumentCommand{\satt}{O{t}}{\operatorname{\sat[#1]}}
\NewDocumentCommand{\satr}{O{\STDRegion}}{\operatorname{\sat[#1]}}

\NewDocumentCommand{\HPlane}{O{t}}{\operatorname{H_{#1}}}

\NewDocumentCommand{\Region}{}{R}
\NewDocumentCommand{\triangulation}{}{\bar\Delta}
\NewDocumentCommand{\prob}{}{\text{Pr}}

\definecolor{r1}{rgb}{0.67, 0.88, 0.69}
\definecolor{r2}{rgb}{0.53, 0.66, 0.42}
\definecolor{r3}{rgb}{0.0, 0.8, 0.6}

\definecolor{darkgreen}{rgb}{0,0.5,0}
\definecolor{backcolour}{rgb}{0.99,0.96,0.89}
\definecolor{keycolour}{rgb}{0.07,0.55,0.84}
\definecolor{valuecolour}{rgb}{0.11,0.63,0.6}
\definecolor{attributecolour}{rgb}{0.57,0.63,0.63}
\definecolor{darkcolour}{rgb}{0.2,0.2,0.2}
\definecolor{orangecolor}{rgb}{1.0,0.6,0.2}

\newif\ifistoreview
\istoreviewtrue

\newcommand{\setreviewsoff}{\istoreviewfalse}
\newcommand{\removeColor}{\textcolor{red}}
\newcommand{\addColor}{\textcolor{blue}}
\newcommand{\remove}[1]{\ifistoreview\protect\removeColor{\st{#1}}\else\unskip\xspace\fi}
\newcommand{\add}[1]{\ifistoreview\protect\addColor{#1}\else #1\fi}
\newcommand{\replace}[2]{\ifistoreview\remove{#1}~\add{#2}\else \remove{#1}\add{#2}\fi}

\DeclareRobustCommand{\hsout}[1]{\texorpdfstring{\st{#1}}{#1}}
\newcommand{\removeSectionTitle}[1]{\ifistoreview\removeColor{\hsout{#1}}\else\unskip\xspace\fi}

\setreviewsoff

\colorlet{BLUE}{blue}
\colorlet{RED}{red}

\input{tikzsettings}

\title{State-space construction of Hybrid Petri nets with multiple stochastic firings}

\author{Jannik H\"uls}
\affiliation{\institution{Westf\"alische Wilhelms-Universit\"at M\"unster}}
\email{jannik.huels@wwu.de}

\author{Carina Pilch}
\affiliation{\institution{Westf\"alische Wilhelms-Universit\"at M\"unster}}
\email{carina.pilch@wwu.de}

\author{Patricia Schinke}
\affiliation{\institution{Westf\"alische Wilhelms-Universit\"at M\"unster}}
\email{patricia.schinke@wwu.de}

\author{Henner Niehaus}
\affiliation{\institution{Westf\"alische Wilhelms-Universit\"at M\"unster}}
\email{henner.niehaus@wwu.de}

\author{Joanna Delicaris}
\affiliation{\institution{Westf\"alische Wilhelms-Universit\"at M\"unster}}
\email{joanna.delicaris@wwu.de}

\author{Anne Remke}
\affiliation{\institution{Westf\"alische Wilhelms-Universit\"at M\"unster}}
\email{anne.remke@wwu.de}

\newcommand{\hypro}{\textsc{HyPro}\xspace}
\newcommand{\hpnmg}{\textsc{hpnmg}\xspace}
\newcommand{\hypeg}{\textsc{hypeg}\xspace}
\newcommand{\ppl}{\textsc{PPL}\xspace}
\newcommand{\polymake}{\textsc{polymake}\xspace}
\newcommand{\cdd}{\textsc{cdd}\xspace}
\newcommand{\cgal}{\textsc{CGAL}\xspace}
\newcommand{\spaceex}{\textsc{SpaceEx}\xspace}

\newcommand{\flowstar}{\textsc{Flow*}\xspace}
\newcommand{\modest}{\textsc{Modest}\xspace}
\newcommand{\stochy}{\textsc{StocHy}\xspace}
\newcommand{\faust}{\textsc{Faust$^2$}\xspace}
\newcommand{\cpntools}{\textsc{CPN Tools}\xspace}
\newcommand{\mobius}{\textsc{M\"obius}\xspace}
\newcommand{\psrecord}{\textsc{psrecord}\xspace}

\date{November 2019}

\acmJournal{TOMACS}
\acmBooktitle{ACM Transactions on Modeling and Computer Simulation (TOMACS)}

\begin{document}

\begin{abstract}
    Hybrid Petri nets have been extended to include general transitions that fire after a randomly distributed amount of time. With a single general one-shot transition the state space and evolution over time can be represented either as a \emph{Parametric Location Tree} or as a \emph{Stochastic Time Diagram}. Recent work has shown that both representations can be combined and then allow multiple stochastic firings. This work presents an algorithm for building the \emph{Parametric Location Tree} with multiple general transition firings and shows how its transient probability distribution can be computed using multi-dimensional integration. We discuss the (dis-)advantages of an interval arithmetic and a geometric approach to compute the areas of integration. Furthermore, we provide details on how to perform a Monte Carlo integration either directly on these intervals or convex polytopes, or after transformation to standard simplices.  A case study on a battery-backup system shows the feasibility of the approach and discusses the performance of the different integration approaches. 
\end{abstract}{}


\maketitle


\input{sections/introduction.tex}
\input{sections/hpng.tex}

\input{sections/plt.tex}
\input{sections/rv.tex}
\input{sections/transient_domains.tex}

\input{sections/transient_geometrical.tex}

\input{sections/casestudy.tex}
\input{sections/conclusion.tex}

\section{\add{Acknowledgements}}
\add{A special thanks goes to Stefan Schupp for his help with the integration of \hypro and performing dedicated measurements on the relevant function calls. Furthermore, we would like to thank the reviewers for their very constructive and positive remarks.}

 \bibliographystyle{ACM-Reference-Format}
 \bibliography{references}

\end{document}

%% file: tikzsettings.tex
\tikzset{
	font=\sffamily,	
	every text node part/.style={align=center},
	cPlace/.style={thick,circle, draw, double distance=0.5mm, minimum size = 1cm},
	dPlace/.style={thick,circle, draw, minimum size = 1.1cm},
	dTrans/.style={thick,rectangle, draw, minimum size = 0.5cm, fill = gray, minimum height = 1.1cm},
	iTrans/.style={thick,rectangle, draw, minimum size = 0.15cm, fill = black, minimum height = 1.1cm},
	gTrans/.style={thick,rectangle, draw, minimum size = 0.5cm, minimum height = 1.1cm},
	cTrans/.style={thick,rectangle, draw, double distance=0.5mm, minimum size = 0.4cm, minimum height = 1cm},
	dyTrans/.style={thick,rectangle, draw, double distance=0.5mm, fill=black, minimum size = 0.4cm, minimum height = 1cm},
	cArc/.style={decoration={markings, mark=at position 1 with {\arrow[scale=1.5]{open triangle 60}}},
		double distance=0.5mm, 
		shorten >= 3.199mm, 
		shorten <= 0.39mm, 
		preaction = {decorate},
		thick},
	dArc/.style={thick,> = stealth', ->, thick},
	gArc/.style={thick,> = triangle 60, <->, thick}, 
	gcArc/.style={thick,> = triangle 60, <->, thick, shorten <= 0.25mm}, 
	gArcNeu/.style={>=spaced triangle 60, <->,bend right=20, very thick},
	gccArc/.style={thick,> = triangle 60, <->, thick, shorten <= 0.25mm, shorten >= 0.25mm}, 
    iArc/.style={thick,-o, thick, shorten <= 0.25mm} 
    }

\tikzstyle{border}=[thick, darkcolour]
\tikzstyle{filling}=[thick, fill, fill opacity = 0.2]
\tikzstyle{cone}=[draw = none, fill, fill opacity = 0.2, text opacity=1]

%% file: sections/introduction.tex
\section{Introduction}

\emph{Hybrid Petri nets with general transitions} (HPnG) \cite{gribaudo2016hybrid} extend Hybrid Petri nets \cite{alla1998continuous} by adding stochastic behaviour through general transitions with a randomly distributed delay. HPnGs provide a high-level formalism for a class of stochastic hybrid systems with piece-wise linear  continuous behaviour without resets and a probabilistic resolution of non-determinism. Hybrid Petri nets have been shown useful for the evaluation of critical infrastructures \cite{ghasemieh2013sewage,jongerden2016does}, even though they have previously been restricted to a single random variable. This paper shows how the state-space of a Hybrid Petri net with multiple  general transition firings can be constructed as a \emph{Parametric Location Tree} (PLT) \cite{gribaudo2016hybrid} for a predefined maximum time. As such, this paper provides the missing link to recent work \cite{hypro_valuetools,huls2019ModelCheckingHPnGs} which provides model checking capabilities for HPnGs with multiple random variables, assuming the existence of the PLT.  

The approach of this paper is twofold:
 Firstly, a purely numerical iterative algorithm for the construction of a PLT in the presence of a finite but arbitrary number of stochastic firings before a certain maximum time is presented. This algorithm is based on introducing an order on the random variables that occur due to the firing of general transitions. Each of these firings affects the evolution of the Petri net. 
 The idea of parametric analysis, as presented in  \cite{gribaudo2016hybrid}, is to collect values of  random variables into intervals, which lead to a similar system evolution. These intervals are called potential domains. Secondly, to facilitate a transient analysis, we identify those parametric locations the HPnG can be  in at a specific point in time (the so-called candidates) and compute those subsets of the domains of the random variables present, for which the HPnG is guaranteed to be in a specific candidate. 
 A multi-dimensional integration over these so-called time-restricted domains yields a transient  distribution over locations. 
 
 The evolution of the state space over time of an HPnG can be partitioned into sets of states with similar behavior. 
 Two approaches exist for representing the time-bounded system executions symbolically, by conditioning their evolution on the firing times of the general transitions:
 One method creates a \emph{Parametric Location Tree} (PLT), which symbolically represents sets of states as  \emph{locations} in the nodes of the tree and events between states as edges between such nodes. The other method uses a geometric representation for sets of states in terms of  convex polytopes (so-called regions) with similar characteristics~\cite{hypro_valuetools}. The idea of a polyhedra-based representation of the state space has been explored before, e.g.,  for (flowpipe) approximations \cite{frehse2013FlowpipeApproximationClustering, frehse2011SpaceExScalableVerification} and to abstract uncountable-state stochastic processes \cite{soudjani2015FAUST2FormalAbstractions,soudjani2013AdaptiveSequentialGridding}.

Both representations are symbolic, having different (dis-)advantages. The PLT-representation can be computed efficiently and lends itself well to compute simple measures like transient distributions. The \add{purely} geometric representation \add{as presented in \cite{godde2017ModelCheckingSTL}} is computationally more expensive, as it heavily relies on hyperplane arrangement, for which to the best of our knowledge no efficient implementations for \replace{higher}{more than three} dimensions exist. 
However, it is possible to compute a geometric representation for each location in a PLT \cite{hypro_valuetools}, that fully relies on halfspace intersection, which can be solved e.g., using \replace{quick hull}{quickhull} \cite{barber1996quickhull} also for higher dimensions. This combines the computational speed of the parametric reachability analysis with the ease \add{that }a geometric representation brings for model checking complex properties \cite{huls2019ModelCheckingHPnGs}. 
 
 The current paper extends the previously published version \cite{huls2019StateSpaceConstructionHybrid}, by a thorough discussion on how the integration can be performed either on the restricted domains or on the equivalent geometric representation as regions which have to be intersected with a time constraint. The resulting arbitrary convex polytopes and their representation need\remove{s} to be adapted for the computation of the corresponding probabilities using multi-dimensional integration. This can be done by performing  triangulation on these convex polytopes, resulting  in a set of simplices. These are then affinely transformed into standard simplices, which allow the use of different multi-dimensional integration techniques. Hence, the change of representation ensures the extensibility of the presented approach to different integration methods.
 
The multi-dimensional integration in this paper is performed using Monte-Carlo methods, either over these standard simplices or directly on convex polytopes to compute transient probabilities. As the integration approach detailed in this paper is applicable to arbitrary convex polytopes, it can also be applied to convex polytopes that form the satisfaction set of an STL formula on a Hybrid Petri net, as described in  \cite{huls2019ModelCheckingHPnGs}.

 
 
  A scalable case study on a battery backed-up system in the presence of short-term peak loads and grid failures shows the feasibility of the approach and compares the performance of the different integration methods, as implemented in the tool \hpnmg~\cite{huels2020hpnmg}.

 \paragraph{Related Work.}

Hybrid Petri nets without general transitions form a subclass of non-initialized singular automata \cite{alla1998continuous}. Hence, unbounded reachability is not decidable and \cite{gribaudo2016hybrid,ghasemieh2012region} resort to computing time-bounded reachability, however only in the presence of at most two random variables. Previous work had to restrict the number of general transition firings, since (i) identifying the potential domains of child locations requires linear optimization,  (ii) transient analysis relies on multi-dimensional integration over polytopes. For one random variable, this was solved using the simplex method and  discretization \cite{gribaudo2016hybrid}. For two random variables, \cite{ghasemieh2012region} used hyperplane arrangement and triangulation to compute numerically exact results.  Vertex enumeration was proposed in \cite{hypro_valuetools} to  construct a PLT for an arbitrary but finite number of stochastic firings, but did not provide a general algorithm. The strict order on the firings of the general transitions,  as in this paper, simplifies the computation of the child locations to solving linear inequalities. This allows the implementation of (i) an efficient iterative algorithm for constructing the PLT up to a certain maximum time and (ii) a transient analysis based on \emph{Fourier-Motzkin} variable elimination.

 The \modest~toolset \cite{hartmanns2014ModestToolsetIntegrated}  provides an extensible and  comprehensive toolset for analytical and statistical model checking. It allows numerous model formalisms and in contrast to the approach presented in this paper, can also deal with non-determinism. Statistical model checking for stochastic hybrid models has been presented in \cite{budde2018StatisticalModelChecker} and for HPnGs in \cite{pilch2017HYPEGStatisticalModel} with various confidence intervals and hypothesis tests. 
 A limited overapproximating model checker for stochastic hybrid automata has also been released as part of \modest~\cite{hahn2013CompositionalModellingAnalysis}.
\spaceex~\cite{frehse2011SpaceExScalableVerification} and \flowstar~\cite{chen2013FlowAnalyzerNonlinear} provide reachability analysis for non-linear hybrid automata  without stochasticity. Also constraint solvers have been proposed for the automated analysis of probabilistic hybrid automata \cite{franzle2010EngineeringConstraintSolvers}. Approaches for hybrid automata extended with discrete probability distributions  \cite{ZHANG2012572,KWIATKOWSKA2002101,sproston2000decidable,TEIGE2009162}  compute time-bounded reachability using abstraction. \add{The tools \faust \cite{soudjani2015FAUST2FormalAbstractions} and \stochy \cite{soudjani2013AdaptiveSequentialGridding} provide model checking capabilities for discrete-time stochastic hybrid models via abstractions to Markov decisision processes. }
While our approach is limited to a restricted class of stochastic hybrid automata, the resulting areas of integration can be computed without approximation or abstraction, both for transient and for reachability analysis. Approximation is only introduced via the multi-dimensional integration, for which an error estimate can be computed.

 Modeling and analysis of reactive timed systems based on various classes of Petri nets has been proposed in \cite{bucci2010ORISToolModeling}. \cpntools~\cite{everdij2008EnhancingHybridState} is a toolset for simulating and analysing Coloured Petri nets (CPNs), for which extensions to Stochastically and Dynamically Coloured Petri nets (SDCPNs) exist \cite{everdij2006HybridPetriNets}. Both approaches allow for quantitative evaluation and qualitative verification for systems with a discrete notion of state. Another  variant of Petri nets with hybrid and stochastic behaviour are Fluid Stochastic Petri nets (FSPNs) \cite{horton1998fluid,gribaudo2001fluid}, which only allow exponentially distributed and immediate discrete transitions. Related Petri net approaches are all restricted w.r.t. the number of continuous variables \cite{horton1998fluid} or to Markovian jumps \cite{everdij2006HybridPetriNets}.  \mobius~\cite{deavours2002MobiusFrameworkIts} supports modeling formalisms like Stochastic Petri Nets (SPNs), Stochastic Automata Networks (SANs) and Markov Chains and mainly implements discrete-event simulation.  HPnGs instead allow for general probability distributions as the analysis abstracts from the specific distribution as long as possible. Furthermore, the number of continuous variables in the model does not restrict the analysis capabilities of our approach. While the approach presented for HPnGs is also general w.r.t. the number of random variables present in the system, its performance clearly suffers with a growing number of dimensions.

Semi-Markov processes have been used to evaluate the dependability of uninterrupted power supply (UPS) systems \cite{yin} and Stochastic Activity networks have been used to study the resilience of smart grid distribution networks \cite{amare}. They are however restricted to negative exponential stochastic behaviour.

\paragraph{Organisation.} 
\add{This paper is further organised as follows.} Section~\ref{formalism} recalls \add{the syntax and semantics of }the modeling formalism, Section~\ref{PLT} repeats the Parametric Location Tree.  Section~\ref{ss:rvs} introduces its construction for an arbitrary but finite number of random variables. Transient analysis via interval arithmetic is explained in Section~\ref{ss:TA}, and Section~\ref{ss:TGeo} introduces a geometric approach for the computation of the areas of integration. Section~\ref{ss:CST} presents a  feasibility study and compares the performance of the different integration approaches, \add{taking into account CPU and memory usage.} Section~\ref{ss:Conc} concludes the paper. 

%% file: sections/hpng.tex
\section{The model formalism of HPnGs}\label{formalism}
HPnGs extend hybrid Petri nets \cite{alla1998continuous} with so called general transitions \cite{gribaudo2016hybrid}.
We present their syntax \add{in Section~\ref{ss:syntax}} and \remove{semantics and give} \add{the model evolution together with} a definition of states and events in \replace{the following}{Section~\ref{ss:model}}.
\add{A small example is provided to illustrate the notation.}

\subsection{\add{Syntax}}\label{ss:syntax} Hybrid Petri nets with multiple general transition firings (HPnG) consist of \remove{the }three components\add{:} \emph{places}, \emph{transitions} and \emph{arcs}, where places are either \emph{discrete} or \emph{continuous}.   A discrete place holds a natural number of tokens and a continuous place contains a non-negative real amount of fluid.
\add{Note that, in the following definition and the remainder of  this paper $\mathbb{N}_0$ indicates the set of natural numbers including zero and $\mathbb{R}^+_0$ indicates the set of non-negative real numbers.
Transitions which change the amount of tokens in discrete places are either \emph{deterministic}, \emph{immediate} or \emph{general}, i.e., they fire after a random delay. Together they are referred to as discrete transitions, as they only change the discrete marking.
This is in contrast to continuous transitions, which only change the continuous marking.}

\noindent\begin{definition}
\label{def:hpnmg}
\add{A \emph{Hybrid Petri net with general transitions} (\emph{HPnG}) is a tuple $(\mathcal{P}, \mathcal{T}, \mathcal{A}, \mathbf{M}_0, \phi)$.} 
\begin{itemize}
    \item \add{$\mathcal{P} = \mathcal{P}^d \cup \mathcal{P}^c$ is a finite set of \emph{places}, partitioned into disjoint sets of \emph{discrete} and \emph{continuous} places.}
Any discrete place $P^d_i \in \mathcal{P}^d$  holds a marking of $m_i \in \mathbb{N}_0$ tokens and any continuous place $P^c_j \in \mathcal{P}^c$ holds a continuous marking described by a fluid level $x_j \in \mathbb{R}^+_0$. 

\item A continuous place has a predefined upper bound, possibly infinity. The lower bound is always zero. 

\item The initial marking $\mathbf{M}_0$ is given by the initial number of tokens $\mathbf{m}_0$ and fluid levels $\mathbf{x}_0$ of all places. 

\item Transitions \add{in the finite set $\mathcal{T}$} change the marking upon firing, i.e., they change the tokens or fluid level of discrete and continuous places. 

\item \add{Places and transitions are connected via arcs in the finite set $\mathcal{A}$. Discrete arcs in $\mathcal{A}^d$ connect discrete places to discrete transitions and vice versa. Correspondingly, continuous arcs in $\mathcal{A}^f$ connect continuous places and continuous transitions. }How the marking is changed is defined by assigning weights and priorities to  \replace{discrete and continuous arcs, i.e., $A
^d \in \mathcal{A}^d$ and $A
^f \in \mathcal{A}^f$ respectively}{those arcs}. 
\replace{These arcs connect places and transitions and d}{D}epending on the direction of connection, \remove{corresponding} places are called input or output places\add{ of transitions}.

\item \remove{General, deterministic and immediate transitions, together also called discrete transitions, change the discrete marking.} A discrete transition \add{in $\mathcal{T}^D \cup \mathcal{T}^I \cup \mathcal{T}^G$ is either deterministic, immediate or general. It} is enabled when its input places match \add{or exceed }the weight of their connecting arcs. \remove{A continuous transition is enabled if all connected input places hold fluid. Guard arcs $A^t \in \mathcal{A}^t$ may further influence the enabling of transitions. They carry a comparison operator and a weight. A discrete guard arc connects a discrete place to any transition and conditions its enabling on the comparison of the discrete marking  and its weight. Correspondingly,  continuous places may be connected to (only)  discrete transitions via  (continuous) guard arcs.} Discrete transitions are each associated with a clock $c_i$, which if enabled evolves with $\frac{dc_i}{dt}=1$, otherwise $\frac{dc_i}{dt} = 0$. Note that, upon disabling, the clock value is preserved. For general transitions this corresponds to the preemptive resume strategy. A deterministic transition $T^D_k \in \mathcal{T}^D$ fires when $c_i$ reaches the predefined transitions firing time. For an immediate transition $T^I_k \in \mathcal{T}^I$ the predefined firing time is always zero. The firing time of a general transition $T^G_m \in \mathcal{T}^G$ is modelled by a cummulative distribution function (CDF), which is assumed to be absolutely continuous. Each \emph{stochastic firing} results in a random variable that follows the CDF of the general transition. \add{By construction, these random variables are independent of each other, since the firing behavior of the general transitions is solely described by its CDF.}

\item \add{A continuous transition is enabled if all connected input places hold fluid and fires at its nominal rate if the connected input and output places are not at either boundary. Every static continuous transition $T^F_n \in \mathcal{T}^{F}$ has a constant nominal flow rate.
Dynamic continuous transitions $T^{Dyn}_o \in \mathcal{T}^{Dyn}$ represent a set $D \subset \mathcal{T}^{F}$ of static continuous transitions. 
Hence their nominal flow rate is a function of the actual flow rates of all static continuous transitions in $D$ \cite{ghasemieh2013sewage}. }
\item 
\add{The rates of continuous transitions that are connected to a continuous place which is at either of its boundaries require \emph{rate adaptation} \cite{gribaudo2016hybrid}, changing the actual flow rate.}
\item \add{Guard arcs $A^t \in \mathcal{A}^t$ may further influence the enabling of transitions. They carry a comparison operator and a weight. A discrete guard arc connects a discrete place to any transition and conditions its enabling on the comparison of the discrete marking  and its weight. Correspondingly,  continuous places may be connected to (only)  discrete transitions via  (continuous) guard arcs. }


\item \add{All parameters are given by the parameter function $\phi$. For details on $\phi$, refer to~\cite{gribaudo2016hybrid}}.
\end{itemize}
\end{definition}

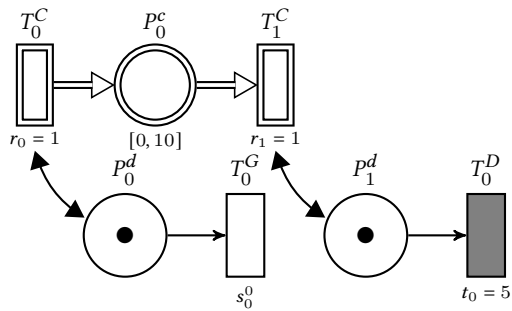
\begin{figure}
    \centering
    \input{images/example_net.tex}
    \caption{\add{Hybrid Petri net with general transition of a reservoir from~\cite{gribaudo2016hybrid}. The continuous place $P^c_0$ models the reservoir, which has a deterministically timed demand of 5 time units, modeled by the deterministic transition $T^D_0$, and a generally distributed pump breaking time, modeled by the general transition $T^G_0$. The initial marking of $P^c_0$ is zero and for the discrete places as shown in the figure. All weights and priorities equal $1$.}}
    \label{fig:example_hpng}
\end{figure}

\add{ With the above definition of arcs, this results in bipartite Hybrid Petri nets, where places form one set and transitions form another set. Furthermore, without guard arcs, the discrete parts of the net and the continuous parts of the net are separate.}
\add{Figure~\ref{fig:example_hpng} presents an example for an HPnG, which models a water reservoir. Places are illustrated by circles, transitions by bars and arcs by arrows.}

\subsection{\add{Model evolution}}\label{ss:model}


\add{Next to the syntax of a hybrid Petri net as defined above, a couple of important concepts exist, which define the evolution of the model over time. These concepts comprise the enabling and firing rules, the sharing of fluid, conflict resolution and rate adaptation. We give a brief overview on these concepts but refer to \cite{gribaudo2016hybrid} for further details.}

 \remove{Every static continuous transition $T^F_n \in \mathcal{T}^{F}$ has a constant nominal flow rate. Dynamic continuous transitions $T^{Dyn}_o \in \mathcal{T}^{Dyn}$ represent a set $D \subset \mathcal{T}^{F}$ of static continuous transitions. Hence their nominal flow rate is a function of the actual flow rates of all static continuous transitions in $D$. Continuous transitions change the fluid level of connected input and output places with a constant rate. The rates of transitions that are connected to a continuous place which is at either of its boundaries require \emph{rate adaptation}, changing the actual flow rate.}

\remove{The rates of transitions that are connected to a continuous place which is at either of its boundaries require \emph{rate adaptation, changing the actual flow rate. At}}\add{For continuous transitions that are connected to places which are at either boundary, a so-called \emph{actual rate} is computed by \emph{rate adaptation}: For places that are at} the upper boundary the inflow \add{of continuous transitions} is decreased to match the outflow\replace{ and}{. For places} at the lower boundary, the outflow of continuous transitions is reduced accordingly. 
A continuous place then evolves with a drift, which equals the sum of the actual inflow rates minus the sum of the actual outflow rates.

\replace{If}{It is possible that} multiple immediate or deterministic transitions \add{are scheduled to} fire at the same time, this \add{so-called} \emph{conflict} is resolved using priorities and weights. For details on \emph{conflict resolution}  and \emph{the concept of enabling}, we refer to \cite{gribaudo2016hybrid}. The probability that a general transition fires at the same \add{point in }time \remove{as}\add{at which} a deterministic \remove{one}\add{transition fires} is zero. We exclude \remove{z}\add{Z}eno behaviour by banning cycles of immediate and  general transitions. Section~\ref{PLT} further describes  the model evolution and the interplay between stochastic and deterministic transitions.

\add{A definition of state is required to formally define the evolution of the model.}

\noindent\begin{definition}
 \label{def:state} \replace{A}{The} state of an HPnG is a tuple $\Gamma = (\mathbf{m}, \mathbf{x}, \mathbf{c}, \mathbf{d}, \mathbf{g}, \mathbf{e})$,  where  $\mathbf{m}$ is the discrete marking, $\mathbf{x}$ the continuous marking,  $\mathbf{c}$ contains for each deterministic transition the enabling time. The drift $\mathbf{d}$ describes the current change of fluid level per time unit for each continuous place,  $\mathbf{g}$ contains the enabling time for the general transitions, and
 $\mathbf{e}$ describes the enabling status of all transitions. 
 
 \end{definition}
 
 Events trigger changes in states as introduced in \cite{gribaudo2016hybrid,pilch2}:

 \noindent\begin{definition}
 \label{def:event} An event 
 $\Upsilon\left(\Gamma_i, \Gamma_{i+1}\right)=\left(\Delta\tau_{i+1}, \varepsilon_{i+1}\right)$ describes the change from one state
 $\Gamma_i$ to another state $\Gamma_{i+1}$, with $\varepsilon_{i+1} \in \mathcal{P}^c\cup\mathcal{T}^I \cup \mathcal{T}^D \cup \mathcal{T}^G \cup\mathcal{A}^t$ specifying the model element that caused the event. Note that,  $\Delta\tau_{i+1} \in \mathbb{R}^+_0$ is a relative time between two events, such that one of the following conditions is fulfilled:
\begin{enumerate}
    \item An immediate, deterministic or general transition fires, such that $\mathbf{m}_i\neq \mathbf{m}_{i+1} \wedge \varepsilon_{i+1} \in \mathcal{T}^I \cup \mathcal{T}^D \cup \mathcal{T}^G $.
    \item A continuous place reaches its lower or upper boundary, such that $\mathbf{d}_i\neq \mathbf{d}_{i+1} \wedge \varepsilon_{i+1} \in \mathcal{P}^c$.
    \item A guard arc condition is fulfilled or stops being fulfilled, such that $\mathbf{e}_i\neq \mathbf{e}_{i+1} \wedge \varepsilon_{i+1} \in \mathcal{A}^t$.
    \end{enumerate}
\end{definition}  


\noindent
The set of all possible events  which can occur in state $\Gamma_i$ is finite and its size depends on the number of continuous places, the number of guard arcs and the number of enabled discrete transitions. It is denoted $\mathcal{E}(\Gamma_i)$ and the set of events with minimum remaining time for that state is defined as follows:
\begin{equation}
\label{eq:setevents}
        \mathcal{E}^\text{min}(\Gamma_i)=\left\{\Upsilon_j(\Gamma_i, \Gamma_j) \in \mathcal{E}(\Gamma_i)\ \middle|\  \nexists \Upsilon_k(\Gamma_i, \Gamma_k)  \in \mathcal{E}(\Gamma_i): \Delta\tau_k < \Delta\tau_j\right\}.
\end{equation}
Note that, multiple events can happen at the same point in time, e.g., due to conflicts between deterministic transitions or due to the scheduling of stochastic transitions. We split the set $ \mathcal{E}^\text{min}(\Gamma_i)$ into two subsets, representing the next random events 
$\mathcal{E}^\text{min}_\text{ran}(\Gamma_i) $ 
and the set of all next deterministic events $\mathcal{E}^\text{min}_\text{det}(\Gamma_i) = \mathcal{E}^\text{min} \setminus \mathcal{E}^\text{min}_\text{ran}(\Gamma_i)$. 
\add{The size of $\mathcal{E}^\text{min}_\text{ran}(\Gamma_i)$ is bounded by the number of enabled general transitions and the size of $\mathcal{E}^\text{min}_\text{det}(\Gamma_i)$ is bounded by the number of continuous places plus the number of guard arcs plus the number of enabled deterministic and immediate transitions.}
\remove{The next minimum event time is unique before the first stochastic firing, it simply is the minimum of the remaining times to fire of all enabled deterministic transitions.}
\add{The time at which the next events may happen can always be determined only depending on the values of the given random variables. The next minimum event time is defined as the minimum time to the next event(s), which can be the same for two or more events.  Also, before the first general transition firing, it is unique and simply is the minimum of the remaining times of all upcoming deterministic events, i.e., deterministic or immediate transition firings, guard arc en-/disablings or continuous boundary events.}
After at least one stochastic firing, the occurrence of events, the clocks and the continuous marking  may linearly depend on the random variable(s). Since an arbitrary but finite number of random variables is present in the system and the computation of the particular next minimum event (for all event types) may depend on all past transition firing times, it is based on polynomials and not scalars.
Hence, after at least one stochastic firing the set of deterministic events with minimum remaining time may consist of more than one element. Their remaining time to fire is derived by minimizing the elements of  $\mathbf{\Delta T}^\text{min}_\text{det}(\Gamma_i)$ over the support of the random variables that already have fired, where
\begin{equation}
\label{eq:setmin}
\mathbf{\Delta T}^\text{min}_\text{det}(\Gamma_i)= \left\{\Delta\tau_m \in \mathbb{R}^+_0\ \middle|\ \exists \Upsilon(\Gamma_i, \Gamma_m) \in \mathcal{E}^\text{min}_\text{det}(\Gamma_i) \right\}.
\end{equation}

One of the main contributions of this paper is to create a Parametric Location Tree (PLT) for multiple general transition firings. The PLT symbolically represents all states, in which an HPnG can be before some predefined maximum time.

%% file: images/example_net.tex
        
 \begin{tikzpicture}[every label/.style={}	, scale=0.8]
 
 \node (tc1) [cTrans, label=above: $T^C_0$, label={below:\footnotesize $r_0=1$}] at (0,0) {};
 \node (pc1) [cPlace, label=above: $P^c_0$, label={below:\footnotesize $[0,10]$}] at (2,0) {};
 \node (tc2) [cTrans, label=above: $T^C_1$, label={below:\footnotesize $r_1=1$}] at (4,0) {};
%

 \node (pd1) [dPlace, label=above: $P^d_1$] at (5.5,-2.5) {\CIRCLE};
 \node (td1) [dTrans, label=above: $T^D_0$, label={below:\footnotesize $t_0=5$}] at (7.5,-2.5) {};
 
 \node (pd2) [dPlace, label=above: $P^d_0$] at (1.5,-2.5) {\CIRCLE};
 \node (tg1) [gTrans, label=above: $T^G_0$, label={below:\footnotesize $s_0^0$}] at (3.5,-2.5) {};

 \draw [cArc] (tc1) to (pc1);
 \draw [cArc] (pc1) to (tc2);
 \draw [dArc] (pd1) to (td1);
\draw [gcArc, bend left=20] (pd1) to ([yshift=-0.45cm]tc2.south);

\draw [dArc] (pd2) to (tg1);
\draw [gcArc, bend left=20] (pd2) to  ([yshift=-0.45cm]tc1.south);

 
%
\end{tikzpicture}
	
	

	

%% file: sections/plt.tex
\section{The Parametric Location Tree}
\label{PLT}

The evolution over time up to a fixed maximum time $\tau_{\text{max}}$ of an HPnG can be described as a so-called Parametric Location Tree (PLT) \cite{gribaudo2016hybrid,hypro_valuetools}. Each node in the tree is called a parametric location and represents a set of \add{potentially uncountably many} states. 

The edges of a PLT represent events. \add{When computing the PLT for a predefined time bound $\tau_\text{max}$, the number of parametric locations is bounded by the number of events, since we exclude Zeno behaviour through the model construction. The PLT then contains by construction all possible execution traces in a symbolic representation, which relies on the random variables modeling the firing times of general transitions. } 
\add{For predefined intervals of the support of random variables (i.e., their domain of definition), every location can be reached at an entry time $\Lambda.t$. }
\add{Choosing an assignment for all random variables from the support of the joint probability distribution function defines a deterministic execution path throughout the PLT. For each time point of this evolution, the system state is unique and is symbolically represented by a specific location.}

\add{Hence, every location contains a symbolic representation of an uncountable set of states, where each of these states can be reached for a specific assignment for all random variables and the evolution time.  This concept also holds for the entry time and the entry state of a location, which can both only be uniquely specified for a fixed assignment of the random variables. Note that both, the support of the random variables and the evolution time are subject to constraints which are computed throughout the construction of the PLT. This is further explained in Section~\ref{ss:rvs}.}

The event corresponding to the edge is called \textit{source event} w.r.t. the child node location. Extending  previous notation, a parametric location is defined as follows:
\begin{definition}
A parametric location $\Lambda$ is defined as a tuple $\Lambda=(\textit{ID}, t, p, \Gamma, \mathbf{S}, \mathbf{o})$, where  the entry time  is  $\Lambda.t$ and  the probability of choosing that specific location with identifier $\Lambda.\textit{ID}$ is $\Lambda.p$ w.r.t. the parent node. This probability $\Lambda.p$ is also denoted as conflict probability, as it equals one if no conflict occurs and in the case of conflicting events, i.e., multiple immediate or multiple deterministic transitions of the same priority are supposed to fire at the same time, the probability is derived from the weights of the conflicting transitions \cite{gribaudo2016hybrid}. $\Lambda.\Gamma$ denotes the  state of the HPnG when entering the location and $\Lambda.\mathbf{S}$ denotes the potential domain of the random variables as described in the following. There can be multiple general transitions present in the system and each can possibly fire multiple times. For each parametric location the firing order that has lead to this location is unique and stored as vector $\Lambda.\mathbf{o}$.
\end{definition}


Each firing of a general transition corresponds to a random variable which equals the enabling time between two consecutive firings. The random variable corresponding to the $j$-th firing of $T^G_i$ is denoted $s_i^j$ and is added to $\mathbf{o}$ upon the firing of the transition. The $r$-th stochastic firing is then stored in $\mathbf{o}[r]$. A new random variable is instantiated each time a general transition becomes enabled for the first time after a previous firing. Hence, the number of random variables $n$ equals the number of stochastic firings $m$ plus the number of general transitions that are currently enabled\add{, thus $m \leq n$}. At a given point in time, a random variable which corresponds to a past firing is said to be \emph{expired}. Accordingly, we define a vector $\mathbf{s}_\Lambda$ of all (expired and not yet expired) random variables for each location $\Lambda$, where\remove{as} the expired random variables are ordered in $\mathbf{s}_\Lambda$ in the same order as specified by $\Lambda.\mathbf{o}$, i.e., $\forall k \in [1,m]: \mathbf{s}_\Lambda[k] = \Lambda.\mathbf{o}[k]$. 
%
Concurrently enabled general transitions yield competing random variables of potentially different distributions, whereas consecutive firings of a single general transition result in a series of identically distributed random variables.  

 The firing of general transitions at different times determine the evolution of the Petri net.  The idea of parametric analysis, as presented in \cite{gribaudo2016hybrid}, is to collect values of  random variables into intervals  $\textbf{S}$ for each random variable $s_i^j$, which lead to the same next event.  These intervals are called potential domains  and need to be specified for each parametric location. Hence the potential domain $\Lambda.\mathbf{S}$  contains a dedicated vector $\mathbf{S}_i$ for each general transition $T^G_i$. Each of these vectors then collects one interval for each  previous firing $j$, stored in  $S_i^j$,  and one additional interval for a potential next firing.
An interval $S_i^j$ has a lower and an upper boundary $S_i^j.l \leq S_i^j.u$, such that location $\Lambda$ is reached if for every $j$, $s_i^j \in [S_i^j.l, S_i^j.u]$. 
This extends the definition presented in \cite{hypro_valuetools} by separately collecting the possible intervals for each general transition. Furthermore, this defines a strict order on the dependencies of each random variable. 
As the number of stochastic firings differs per location, the size of the potential domain, as indicated by $|\mathbf{S}|$, also differs.

All parametric locations, i.e., all symbolic representations of states, of an HPnG are collected in a so-called parametric location tree as defined in the following:

\begin{definition}
The parametric location tree (PLT) of an HPnG is defined as a tree $(\mathbf{V}, \mathbf{E}, v_{\Lambda_0})$, where $\mathbf{V}$ is the set of nodes representing the parametric locations of the HPnG. $\mathbf{E}$ is the set of edges with $e_i = (v_{\Lambda_j}, v_{\Lambda_k}) \in \mathbf{E}$ for $v_{\Lambda_j}, v_{\Lambda_k} \in \mathbf{V}$ if an event $\Upsilon(\Lambda_j.\Gamma, \Lambda_k.\Gamma)$ exists which leads from the parametric location $\Lambda_j$ to its child location $\Lambda_k$.  The root node $v_{\Lambda_0}$  represents the initial location, initialized as follows: $\Lambda_0 = (0, 0, 1, \Gamma_0, \textbf{S}^\infty, \mathbf{o}_0)$, where $\textbf{S}^\infty$ denotes the not yet limited domain for all currently enabled random variables. Since initially no general transitions has fired, the vector $\mathbf{o}_0$ is empty.
\end{definition}

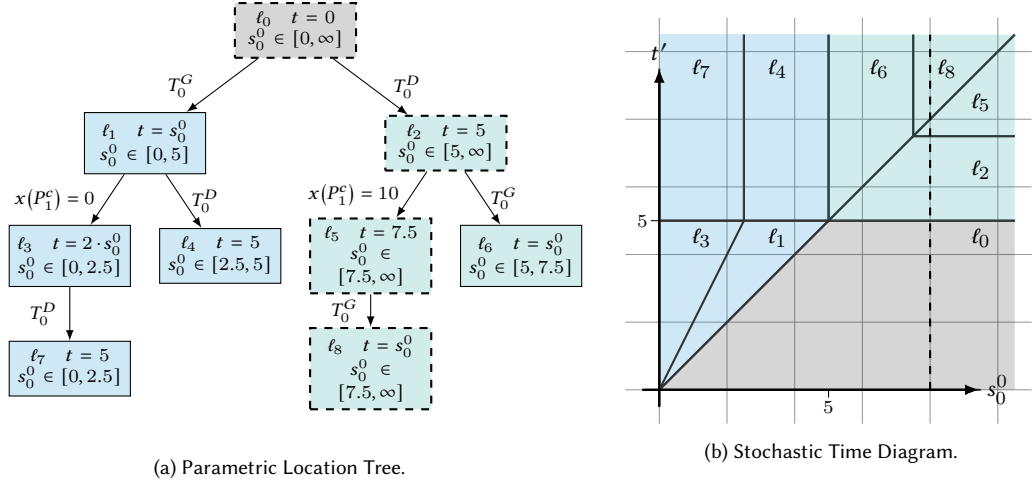
\begin{figure}
    \centering
    \begin{subfigure}[c]{0.48\textwidth}
        \centering
        \input{images/example_plt}
        \subcaption{\add{Parametric Location Tree.}}
        \label{fig:example_plt_std_a}
    \end{subfigure}
    \begin{subfigure}[c]{0.48\textwidth}
        \centering
        \input{images/example_std}
        \subcaption{\add{Stochastic Time Diagram.}}
        \label{fig:example_plt_std_b}
    \end{subfigure}
    \caption{\add{Parametric Location Tree and geometric representation as Stochastic Time Diagram (\cite{ghasemieh2012region}) corresponding to the HPnG in Figure~\ref{fig:example_hpng}. For each location $\Lambda$ the illustration of the PLT contains the \textit{ID}, the entry time $\Lambda.t$ and the domain of the random variable $s_0^0$. The Stochastic Time Diagram illustrates the relation of the possible values for $s_0^0$ and the global time $t'$ for each location.
    Choosing a specific assignment for the random variable, i.e., $s_0^0=8$, leads to a unique path which corresponds to the dashed line indicated in \subref{fig:example_plt_std_b}. The parametric locations which are passed with this specific evolution are highlighted with a dashed border in the PLT representation in \subref{fig:example_plt_std_a}. 
     }
     }
    \label{fig:example_plt_std}
\end{figure}
 
The PLT is iteratively constructed by adding a child location $\Lambda_c$ for each possible event that can take place from a given parametric location. 
The finite set of possible next events 
(\add{see}\remove{c.f.} Equation \ref{eq:setevents}) together with the exclusion of cycles of immediate and general transitions in the model definition yields a finite PLT if computed up to time $\tau_\text{max}$. 

The absolute point in time at which the event takes place is then stored as the locations entry time $\Lambda_c.t$. The other parameters of the location are derived by executing the event in the HPnG. The random variable domain for a specific child location is derived from the parent location, by taking into account the event that leads to the specific location. 
Since the location entry times may depend on previously expired random variables,  the set which consists of the remaining minimum times for all deterministic events $\mathbf{\Delta T}^\text{min}_\text{det}(\Lambda_p.\Gamma)$ may contain more than one element and the order of the timed events depends on the values of the random variables that have expired.  Since different child nodes symbolically describe  sets of values leading to different system evolutions, the resulting potential domains are always disjoint for non-conflicting successors.


\add{Figure~\ref{fig:example_plt_std} shows the resulting state space for the reservoir example from Figure~\ref{fig:example_hpng}.
Figure~\ref{fig:example_plt_std_a} shows the PLT and Figure~\ref{fig:example_plt_std_b} provides a geometrical represenation of the same state-space, i.e., the  Stochastic Time Diagram~\cite{ghasemieh2012region}, which allows to illustrate a unique path which corresponds to the evolution of time for a given assignment of the random variable, e.g. the dashed vertical line that starts for $s_0^0=8$ and crosses through locations $\ell_0, \ell_2, \ell_5$ and $ \ell_8$. 
}






%% file: images/example_plt.tex
\begin{tikzpicture}[font=\footnotesize,loc/.style={rectangle, text width=1.4cm, draw, text opacity = 1}, scale=1, pltfilling/.style={fill opacity = 0.2}]
	\node[loc, pltfilling, fill=darkcolour, dashed, thick](0) at (0,0) {$\ell_0 \quad t=0$\\ $s_0^0 \in [0,\infty]$};
	
	\node[loc, pltfilling, fill=keycolour](1) at (-2,-1.5) {$\ell_1 \quad t=s_0^0$\\ $s_0^0 \in [0,5]$};
	
	\node[loc, pltfilling, fill=valuecolour, dashed, thick](2) at (2,-1.5) {$\ell_2 \quad t=5$\\ $s_0^0 \in [5,\infty]$};
	
	\node[loc, pltfilling, fill=keycolour](3) at (-3,-3) {$\ell_3 \quad t=2\cdot s_0^0$\\ $s_0^0 \in [0,2.5]$};
	
	\node[loc, pltfilling, fill=keycolour](4) at (-1,-3) {$\ell_4 \quad t=5$\\ $s_0^0 \in [2.5,5]$};
	
	\node[loc, pltfilling, fill=valuecolour, dashed, thick](5) at (1,-3) {$\ell_5 \quad t=7.5$\\ $s_0^0 \in [7.5,\infty]$};
	
	\node[loc, pltfilling, fill=valuecolour](6) at (3,-3) {$\ell_6 \quad t=s_0^0$\\ $s_0^0 \in [5,7.5]$};
	
	\node[loc, pltfilling, fill=keycolour](7) at (-3,-4.5) {$\ell_7 \quad t=5$\\ $s_0^0 \in [0,2.5]$};
	
	\node[loc, pltfilling, fill=valuecolour, dashed, thick](8) at (1,-4.5) {$\ell_8 \quad t=s_0^0$\\ $s_0^0 \in [7.5,\infty]$};
	
	\draw[-latex] (0) to node[left, inner sep=3mm] {$T^G_0$} (1);
	\draw[-latex] (0) to node[right, inner sep=3mm] {$T^D_0$} (2);
	\draw[-latex] (1) to node[left, inner sep=1mm] {$x\bigl({P_1^c}\bigr)=0$} (3);
	\draw[-latex] (1) to node[right, inner sep=1mm] {$T^D_0$} (4);
	\draw[-latex] (2) to node[left, inner sep=1mm] {$x\bigl({P_1^c}\bigr)=10$} (5);
	\draw[-latex] (2) to node[right, inner sep=1mm] {$T^G_0$} (6);
	\draw[-latex] (3) to node[left] {$T^D_0$} (7);
	\draw[-latex] (5) to node[left] {$T^G_0$} (8);
	
	\end{tikzpicture}

%% file: images/example_std.tex
 \begin{tikzpicture}[scale=0.9]
 \useasboundingbox (-0.5,-0.5) rectangle (5.5,5.5);
 \draw[step=1.0, gray!80!white,thin] (-0.5,-0.5) grid (5.5,5.5);
	\node (ursprung) at (0,0) {};
	\node[] (y) at (0,5) {$t'$};
	\node[] (x) at (5,0) {$s_0^0$};
	\draw[-latex,thick] ($(ursprung)-(0,0.25)$) to (y);
	\draw[-latex,thick] ($(ursprung)-(0.25,0)$) to (x);
	\node[inner sep=1] (t5) at (-0.25,2.5) {\footnotesize $5$};	
	\draw (0,2.5) -- (t5);
	\node[inner sep=1] (s5) at (2.5, -0.25) {\footnotesize $5$};
	\draw (2.5,0) -- (s5);

\draw[border] (5.25,2.5) -- (2.5,2.5) -- (0,0) -- (4.75,0);
\draw[cone, fill=darkcolour] (5.25,2.5) -- (2.5,2.5) -- (0,0) -- (5.25,0) -- cycle;
\node (l0) at (4.75,2.25) {$\ell_0$};

\draw[border] (5.25,2.5) -- (2.5,2.5) -- (3.75,3.75) -- (5.25,3.75);
\draw[cone, fill=valuecolour]  (5.25,3.75)  -- (5.25,2.5) -- (2.5,2.5)-- (3.75,3.75) -- cycle;
\node (l0) at (4.75,3.25) {$\ell_2$};

\draw[border] (5.25,3.75) -- (3.75,3.75) -- (5.25,5.25);
\draw[cone, fill=valuecolour] (5.25,5.25) -- (3.75,3.75) --(5.25,3.75) --  cycle;
\node (l0) at (4.75,4.25) {$\ell_5$};

\draw[border] (5.25,5.25) -- (3.75,3.75) -- (3.75,5.25);
\draw[cone, fill=valuecolour] (5.25,5.25) -- (3.75,3.75) -- (3.75,5.25) -- cycle;
\node (l0) at (4.25,4.75) {$\ell_8$};

\draw[border] (2.5,5.25) -- (2.5,2.5) -- (3.75,3.75) -- (3.75,5.25);
\draw[cone, fill=valuecolour] (3.75,5.25) -- (2.5,5.25) -- (2.5,2.5) -- (3.75,3.75) --  cycle;
\node (l0) at (3.25,4.75) {$\ell_6$};

\draw[border] (2.5,5.25) -- (2.5,2.5) -- (1.25,2.5) -- (1.25,5.25);
\draw[cone, fill=keycolour] (2.5,5.25) -- (2.5,2.5) -- (1.25,2.5) -- (1.25,5.25) -- cycle;
\node (l0) at (1.75,4.75) {$\ell_4$};

\draw[border] (0,4.75) -- (0,2.5) -- (1.25,2.5) -- (1.25,5.25);
\draw[cone, fill=keycolour] (1.25,5.25) -- (0,5.25) -- (0,2.5) -- (1.25,2.5) --  cycle;
\node (l0) at (0.625,4.75) {$\ell_7$};

\draw[cone, fill=keycolour] (0,0) -- (0,2.5) -- (1.25,2.5) --  cycle;
\node (l0) at (0.625,2.25) {$\ell_3$};
\draw[border] (0,0) -- (0,2.5) -- (1.25,2.5);

\draw[cone, fill=keycolour] (0,0) -- (2.5,2.5) -- (1.25,2.5) --  cycle;
\node (l0) at (1.75,2.25) {$\ell_1$};
\draw[border] (0,0) -- (2.5,2.5) -- (1.25,2.5) -- (0,0);


\draw[dashed, thick] (4,0) -- (4,5.25);

 \draw[-latex,thick] ($(ursprung)-(0,0.25)$) to (y);
 \draw[-latex,thick] ($(ursprung)-(0.25,0)$) to (x);
\end{tikzpicture}	 

%% file: sections/rv.tex
\section{\add{PLT construction}\removeSectionTitle{Random Variables Support}}\label{ss:rvs}

When allowing more than one stochastic firing the construction of the PLT becomes more complex \cite{hypro_valuetools}. For each enabled general transition, at least one location is added as child node, where the edge represents its next firing before every other event. Additionally, each general transition needs to be scheduled after each minimum deterministic event. Hence, if there are $n$ deterministic events, for each enabled general transition, first $n+1$ child nodes are generated, with one edge for the general transition firing and $n$ for all possible deterministic next events $\mathcal{E}^\text{min}_\text{det}(\Gamma_i)$.  The deterministic next events and the enabled general transitions can then be arranged in two different ways: (i) either deterministic events are scheduled first (ii) or a general transitions fires before the next deterministic events. We need to consider all possible combinations of deterministic events and general transitions firings and we need to adjust the boundaries of the potential domains in all cases accordingly, which is discussed in the following.

\add{First, Section~\ref{ss:expiredrv} explains how the intervals of expired random variables are limited, such that the locations for the deterministic events, whose occurrence time might depend on those random variables, can be correctly scheduled. Section~\ref{ss:childlocations} then describes the scheduling of the  child locations in detail.}

\subsection{Adjust boundaries for expired random variables} \label{ss:expiredrv}
\replace{Each parametric location provides an interval $\mathbf{S}_i^j$ of}{The support of each random variable is partitioned into the potential domains as explained in Section~\ref{PLT}. These intervals $\mathbf{S}_i^j$ contain} possible values for every random variables $s_i^j$ \replace{, as previously described}{ per location}. The bounds of those intervals  may depend on firings with a lower order. When scheduling the deterministic successors of a  location, we first need to compute the set of minimum events. In case this set contains more than one element, their order depends on the values of other random variables.  This is handled by adjusting the potential domain for the corresponding random variables per event, such that the event takes place before all other minimum events for all values in the (adjusted) potential domain of the corresponding successor location.

The computation of the correct potential domain corresponds to minimizing the time to the minimum event over the domain of all random variables. In earlier work this has been solved, e.g., using the simplex method \cite{gribaudo2016hybrid} in case of one random variable, hyperplane arrangement in the case of two \cite{ghasemieh2012region} or as suggested by \cite{hypro_valuetools} using vertex enumeration. Instead, this paper defines a strict total order $\prec$ on the random variables that is based on their firing order: Let $s_i^j$ and $s_k^l$ be two random variables, then $s_i^j \prec s_k^l$ holds iff the $i$-th general transition fires for the $j$-th time before the $k$-th general transition fires for the $l$-th time: 
 \begin{equation}\exists u, v \in \mathbb{N}: \mathbf{o}[u] = s_i^j \wedge \mathbf{o}[v] = s_k^l \wedge u< v \Rightarrow s_i^j \prec s_k^l.\end{equation}
 This ensures that the interval bounds of a random variable $s_k^l$ may only depend on earlier expired random variables $s_i^j$. The interval $S_k^l$ of a random variable  $s_k^l$ can in general be described by two  linear expressions:
\begin{equation}
\label{eq:lin_eq}
s_k^l\in S_k^l = \left[S_k^l.l, S_k^l.u \right] = \left[ a_0 + \sum\limits_{s_i^j\prec s_k^l} a_i^j \cdot s_i^j,\ b_0 + \sum\limits_{s_i^j \prec s_k^l} b_i^j \cdot s_i^j \right], \text{ with } a_0, a_i^j, b_0, b_i^j \in \mathbb{R}. 
\end{equation}
\noindent
Each interval bound is given by a real constant plus the sum of previously expired random variables, each potentially multiplied by a real constant. In the following we describe how these bounds are computed.

 \replace{Scheduling}{To schedule} the deterministic successors of a  parametric location $\Lambda_p$,  the set of minimum events $\mathcal{E}^\text{min}_\text{det}(\Lambda_p.\Gamma)$ is required. 
 When generating a child location $\Lambda_c$ for a successor that results from a deterministic event  $\Upsilon(\Lambda_p.\Gamma, \Lambda_c.\Gamma)  \in \mathcal{E}^\text{min}_\text{det}(\Lambda_p.\Gamma)$, its potential domain is determined as follows: In a first step, the potential domain of $\Lambda_c$ is set to the one of $\Lambda_p$. 
 If the set $\mathbf{\Delta T}^\text{min}_\text{det}(\Lambda_p.\Gamma)$  contains more than one value, the intervals for the expired random variables in the potential domain $\Lambda_c.\mathbf{S}$ have to be limited for the corresponding deterministic successor, such that its source event $\Upsilon(\Lambda_p.\Gamma,\Lambda_c.\Gamma)=(\Delta \tau_c,\varepsilon_c)$, takes place before any other event. Thus, $\Delta \tau_c$ is required to be smaller than or equal to the remaining time  for any other event:
\begin{equation}
\label{eq:times}
\forall\Delta \tau^*\in \mathbf{\Delta T}^\text{min}_\text{det}(\Lambda_p.\Gamma): \Delta \tau_c\leq\Delta \tau^*.
\end{equation}


Hence, the potential domain of the random variables in $\Lambda_c$ is adjusted by comparing $\Delta \tau_c$  in a pairwise fashion to each other value $\Delta \tau^*\in \mathbf{\Delta T}^\text{min}_\text{det}(\Lambda_p.\Gamma)$ and limiting the intervals of the corresponding random variables such that  Equation \ref{eq:times} holds for the affected potential domains. Having $\Delta \tau_c = \Delta \tau^*$ means that both linear expressions intersect. Note that, this results in closed intervals for the potential domains with overlapping interval bounds, which however does not make a difference, since the probability for such case equals zero.\add{\footnote{\add{Working with only closed intervals  facilitates the technical representation of the intervals in the implementation.}}}

The computation can be done by considering only  random variables which correspond to firings in the past, i.e., with a lower order, since only past firings can affect clocks and markings and the remaining time to fire, as stored in  $\Delta \mathbf{T}^\text{min}_\text{det}(\Lambda_p.\Gamma)$.
Since $\Delta \tau_c$ is compared to the other remaining times one-by-one, we first present the pairwise comparison in the following and then explain how the results are brought together to obtain the potential domain for the child location $\Lambda_c$.
The remaining times to fire, $\Delta \tau_c$ and each $\Delta \tau^* \in \mathbf{\Delta  T}^\text{min}_\text{det}(\Lambda_p.\Gamma)$ for $\alpha_0, \alpha_z, \beta_0, \beta_z \in \mathbb{R}$ and $m = \vert \mathbf{o} \vert$, can be written as:
 \begin{equation}
 \label{eq:remaining_time}
 \Delta \tau_c=\alpha_0+\sum^m_{z=1}\alpha_z \cdot \mathbf{o}[z], 
\text{ and } \Delta \tau^*=\beta_0+\sum^m_{z=1}\beta_z \cdot \mathbf{o}[z].
\end{equation}


\add{Note that, the coefficients $\alpha_i$ follow from the computation of the source event leading to this parametric location $\Lambda_c$. Specifically, the time to this event $\Delta \tau_c$ is computed according to the event type, defined in Section~\ref{formalism}. Accordingly, the coefficients $\beta_i$ follow from the events in the set $\mathcal{E}^\text{min}_\text{det}(\Lambda_p.\Gamma)$.}

According to Equation \ref{eq:lin_eq}, the expressions in Equation \ref{eq:remaining_time} are also dependent on the random variables in firing order. Each variable may again be multiplied by a real constant, whose computation is presented in the following.  
The interval for any random variable $\mathbf{o}[z]$ with the same multiplier in both linear expressions, i.e., $\alpha_z = \beta_z$, does not need to be adapted since the validity of $\Delta \tau_c\leq\Delta \tau^*$ is independent of this random variable. 

Using these linear expressions, the inequality  $\Delta \tau_c\leq\Delta \tau^*$ can be solved for a specific random variable, i.e., the random variable with the highest order for which the multipliers differ in both inequalities. Hence, to ensure that the condition $\Delta \tau_c\leq\Delta \tau^*$ is fulfilled, it is sufficient to limit the intervals of the potential domain of this specific random variable, such that the rearranged inequality (that is solved for this random variable) is fulfilled. 
Precisely, we determine the maximum index $k$ with respect to the order $\prec$, for which $\alpha_k\neq \beta_k$ holds.
  If no such index $k$ exists, it follows that $\Delta \tau_c=\Delta \tau^*$, which indicates two events with the same remaining occurrence time. In this case  only the event with the higher order (\replace{c.f.}{see} Table 1 in \cite{gribaudo2016hybrid}) is considered. If both events have the same order, the conflict resolution (\replace{c.f.}{see} Section~3.4 in \cite{gribaudo2016hybrid}) is used to compute  probabilities and $\Lambda_c.p$ is updated accordingly and the potential domain does not need to be adjusted. Due to the definition of $\mathcal{E}^\text{min}_\text{det}(\Gamma_i)$, if such a maximum index $k$ exists, it needs to be larger than zero. Otherwise, the linear expressions of the remaining times $\Delta \tau_c$ and $\Delta \tau^*$ would be parallel functions and one of the remaining event times is always larger and hence the corresponding event is not included in  $\mathcal{E}^\text{min}_\text{det}(\Gamma_i)$. For any $k>0$, the potential domain of the random variable  $\mathbf{o}[k]$ is adjusted by rearranging the inequality $\Delta \tau_c\leq\Delta \tau^*$ as: 
  \begin{equation}
 \alpha_0+\sum\limits_{z=1}^m \alpha_z \cdot \mathbf{o}[z] \leq \beta_0+\sum\limits_{z=1}^m \beta_z \cdot \mathbf{o}[z]
  \Leftrightarrow(\alpha_k-\beta_k)\cdot \mathbf{o}[k]\leq (\beta_0-\alpha_0) + \sum\limits_{z=1}^{k-1} (\beta_z-\alpha_z)\cdot \mathbf{o}[z]. 
 \end{equation}
\noindent
First $\Delta \tau_c$ as well as $\Delta \tau^*$ are replaced by the expressions presented in Equation \ref{eq:remaining_time}. Then both summations are combined on the right side of the inequality and the largest differentiating term, i.e., the one with index $k$, is pushed to the left side of the inequality. The sum $\sum\limits_{z=k+1}^m (\beta_z-\alpha_z)\cdot \mathbf{o}[z]$ can be omitted since $\forall z>k:$ $\beta_z-\alpha_z= 0$. 
\add{The adjusted potential domain of $\mathbf{o}[k]$ is obtained by }\replace{R}{r}earranging the inequality and  dividing it  by  $\alpha_k-\beta_k$, which is always possible as  $\alpha_k-\beta_k \neq 0$ holds by definition of the index $k$.  We need to distinguish two cases: 

 \emph{Case $\alpha_k>\beta_k$}:
 	\vspace*{-0.5cm}
 	\begin{equation}
 	\begin{split}
 	 (\alpha_k-\beta_k)\cdot \mathbf{o}[k] &\leq (\beta_0-\alpha_0) + \sum\limits_{z=1}^{k-1}(\beta_z-\alpha_z)\cdot \mathbf{o}[z] \\
 	\overset{\alpha_k>\beta_k}{\Rightarrow}\quad\quad\quad  \mathbf{o}[k]&\leq \frac{\beta_0-\alpha_0}{\alpha_k-\beta_k} + \sum\limits_{z=1}^{k-1}\frac{\beta_z-\alpha_z}{\alpha_k-\beta_k} \cdot \mathbf{o}[z].
 	 	\end{split}
 	\end{equation}
 The upper bound for the random variable $s_i^j$ stored in $\mathbf{o}[k]$ is adjusted as follows: 
 	\begin{align}
 	 \Lambda_c.S_i^j = \left[\Lambda_c.S_i^j.l, \frac{\beta_0-\alpha_0}{\alpha_k-\beta_k} + \sum\limits_{z=1}^{k-1}\frac{\beta_z-\alpha_z}{\alpha_k-\beta_k}\cdot \mathbf{o}[z]\right].
 	\end{align}
 	
 	\emph{Case $\alpha_k<\beta_k$}: 
 	\vspace*{-0.5cm}
 	\begin{equation}
 	\begin{split}
 	\quad (\alpha_k-\beta_k)\cdot \mathbf{o}[k] &\leq (\beta_0-\alpha_0) + \sum\limits_{z=1}^{k-1}(\beta_z-\alpha_z)\cdot \mathbf{o}[z] \\
 	\overset{\alpha_k<\beta_k}{\Rightarrow}\quad\quad\quad \mathbf{o}[k]&\geq \frac{\beta_0-\alpha_0}{\alpha_k-\beta_k} + \sum\limits_{z=1}^{k-1}\frac{\beta_z-\alpha_z}{\alpha_k-\beta_k} \cdot \mathbf{o}[z].
	\end{split}
	\end{equation}
The lower bound for the random variable $s_i^j$ stored in $\mathbf{o}[k]$ is adjusted as follows:
	\begin{align}
 	\Lambda_c.S_i^j = \left[\frac{\beta_0-\alpha_0}{\alpha_k-\beta_k} + \sum\limits_{z=1}^{k-1}\frac{\beta_z-\alpha_z}{\alpha_k-\beta_k}\cdot \mathbf{o}[z],\Lambda_c.S_i^j.u \right].
	\end{align}


Repeating the pairwise comparison for every $\Delta \tau^*\in \mathbf{\Delta T}^\text{min}_\text{det}(\Lambda_p.\Gamma)$, we compute the resulting upper and lower bounds and update the potential domain for child $\Lambda_c$ by taking the maximum of the pair-wisely computed lower bounds and the minimum of the corresponding upper bounds. 
This ensures that the resulting potential domain for each random variable equals the intersection of all intervals obtained by the pairwise comparison of $\Delta \tau_c$ and any $\Delta \tau^*\in \mathbf{\Delta T}^\text{min}_\text{det}(\Lambda_p.\Gamma)$. 

For each scheduled child location $\Lambda_c$, the potential domain further needs to be adjusted in case one or more general transitions are currently enabled in the parent location. This is discussed in the following section, where we assume that for all expired random variables, the intervals in $\Lambda_{c'}.\mathbf{S}$ equal the corresponding intervals in $\Lambda_c.\mathbf{S}$, that have been determined in the procedure described above.

\subsection{Scheduling the child locations}
\label{ss:childlocations}
After computing the potential domains for each minimum next event, the corresponding child location needs to be scheduled for each combination of possible next deterministic event and each enabled stochastic firing. This requires iterating over all elements in the set of minimum deterministic events, generating the corresponding child locations and computing the potential domain,   as explained in the previous section. Then the firings of all enabled general transitions need to be scheduled before each minimum next event and the corresponding child locations are generated. To ensure that all random variables are scheduled once before and once after each minimum next event,   the intervals of the currently enabled random variables in the potential domain need to be further restricted. 

\begin{algorithm}[tb]
	\small \ttfamily
	\begin{algorithmic}[1]
		\caption{scheduling child locations}
		\label{alg:scheduling}
	    \FORALL {(deterministicEvent $\Upsilon$ : $\mathcal{E}^\text{min}_\text{det}(\Lambda_p.\Gamma)$)}
	        \STATE $\Lambda_c$ := generateChild(\add{$\Lambda_p$, $\Upsilon$});
	         \STATE $\Lambda_{c}$.setExpiredRVBounds($\Upsilon$);
	        \FORALL {(\replace{s: enabled RV}{stochasticEvent $\Upsilon_s$ : $\mathcal{E}^\text{min}_\text{ran}(\Lambda_p.\Gamma)$})}
	            \STATE $\Lambda_c$.s.l = $\Lambda_p$.s.l + $\deltat_c$;
	            \COMMENT{Eq~\ref{eq:inlowbound}}
	            \STATE $\Lambda_{c'}$ := generateChild(\add{$\Lambda_p$, $\Upsilon_s$});
	            \STATE $\Lambda_{c'}$.copyExpiredRVBounds($\Lambda_c$);
	            \STATE $\Lambda_{c'}$.s.u = $\Lambda_p$.s.l + $\deltat_{c}$;
	            \COMMENT{Eq.~\ref{eq:dehighbound}}
	            \FORALL {(\replace{s' : enabled RV}{stochasticEvent $\Upsilon_{s'}$ : $\mathcal{E}^\text{min}_\text{ran}(\Lambda_p.\Gamma)$})}
	                \IF{(s != s')}
	                    \STATE $\Lambda_{c'}$.s'.l = $\Lambda_p$.s'.l + s - $\Lambda_p$.s.l; 
	                    \COMMENT{Eq.~\ref{eq:inlowforall}}
	               \ENDIF
	           \ENDFOR
	       \ENDFOR
	   \ENDFOR
	\end{algorithmic}
\end{algorithm}


Algorithm \ref{alg:scheduling} presents  pseudo code for the scheduling of all child locations for a given parametric location $\Lambda_p$. In line~2 the function \texttt{generateChild(\add{$\Lambda_p, \Upsilon$})} creates a location for each child  $\Lambda_c$ \add{of parent $\Lambda_p$} for every event \add{$\Upsilon$} in the set of minimum deterministic events $\mathcal{E}^\text{min}_\text{det}$.  The entry time of each child location $\Lambda_c.t$ is set to $\Lambda_p.t + \Delta \tau_c$ and the interval bounds for expired random variables are adjusted as explained in the previous section by calling the function \texttt{$\Lambda_{c}$.setExpiredRVBounds}($\Upsilon$), where $\Upsilon$ is the source event of $\Lambda_c$ (line~3).
In order to schedule  all competing general transition firings after the event $\Upsilon(\Lambda_p.\Gamma, \Lambda_c.\Gamma)  = (\Delta \tau_c, \varepsilon_c)$, \add{all stochastic events $\Upsilon_s \in \mathcal{E}^\text{min}_\text{ran}(\Lambda_p.\Gamma)$ are iterated (line 4) and} the lower interval bound for each of the still enabled random variables is increased by $\Delta \tau_c$ in the location $\Lambda_c$ (lines~ 5--6). Formally, for the continuously enabled random variable $s_j^j$, the lower interval bound $\Lambda_c.S_i^j.l $ of the potential domain $\Lambda_c.\mathbf{S}$ is adapted by increasing the lower bound  $\Lambda_p.S_i^j.l$ of the potential domain of the parent location:
\begin{equation}\label{eq:inlowbound}
\Lambda_c.S_i^j.l =  \Lambda_p.S_i^j.l  + \Delta \tau_c.
\end{equation}
After adapting the already scheduled successors which correspond to deterministic events, we need to 
 schedule each enabled stochastic firing before each of the deterministic events. This requires the construction of new parametric locations. Hence, for each enabled stochastic firing, i.e., for each event $\Upsilon(\Lambda_p.\Gamma, \Lambda_{c'}.\Gamma)  = (\Delta \tau_{c'}, \varepsilon_{c'})  \in \mathcal{E}^\text{min}_\text{ran}(\Lambda_p.\Gamma)$ one \remove{or more } new parametric location\remove{s} $\Lambda_{c'}$ \replace{are}{is} constructed (line \remove{7}\add{6}) \add{for each instance of the for loop. Furthermore,} \remove{and} the intervals of the expired random variables are inherited from the corresponding deterministic location $\Lambda_{c}$ by calling the function \texttt{$\Lambda_{c'}$.copyExpiredRVBounds}($\Lambda_c$) (line~\remove{8}\add{7}).

Furthermore, the upper interval bound of the random variable $s_i^j$ that related to the stochastic firing is decreased in the new child location $\Lambda_{c'}$, such that the firing takes place before any other competing event occurs (line~\remove{9-10}\add{8}). Consequently, $s_i^j$ cannot take any value that is larger than the sum of its previous \replace{upper}{lower} bound 
in the parent location $\Lambda_p$ and the time to the next competing event. 
The upper interval bound  $\Lambda_{c'}.S_i^j.u$ of the child location is decreased to ensure that the enabled random variable $s_i^j$ fires first:
\begin{equation}\label{eq:dehighbound}
\Lambda_{c'}.S_i^j.u =  \Lambda_p.S_i^j.l  + \Delta \tau_c.
\end{equation}



Next, the potential domains for all other enabled random variables have to be adjusted, such that the corresponding firings do not occur before the firing related to $s_i^j$ takes place. Hence, for all enabled random variables (except $s_i^j$), the potential domain is adjusted (lines~\remove{11-16}\add{9--11}). For every random variable $s_u^v \neq s_i^j$ that corresponds to the $v$-th firing of transition $T_u^G \in \mathcal{T}^G$, which is enabled in $\Lambda_p.\Gamma$, the lower interval bound $\Lambda_{c'}.S_u^v.l$ is increased:

\begin{equation}\label{eq:inlowforall}
\Lambda_{c'}.S_u^v.l =  \Lambda_p.S_u^v.l  + s_i^j - \Lambda_p.S_i^j.l.
\end{equation}

Furthermore, the support of the random variable $s_i^{j+1}$ that corresponds to the next firing of $T_i^G$ has to be initialized if $T_i^G$ is still enabled in $\Lambda_{c'}.\Gamma$. The interval is then always set to $\Lambda_{c'}.S_i^{j+1}=[0,\infty)$. All of the new child locations $\Lambda_c$ and $\Lambda_{c'}$ together create the set of children of location $\Lambda_p$, i.e., $\text{children}(\Lambda_p).$

The computational complexity of Algorithm \ref{alg:scheduling} follows the nested \emph{for} loops and is in $\mathcal{O}\Bigl(n^2 \times \left\vert \mathcal{E}^\text{min}_\text{det}(\Lambda_p.\Gamma)  \right\vert\Bigr)$, where $n = \vert \mathbf{S} \vert$ forms an upper bound on the number of enabled random variables.  The size of $\mathcal{E}^\text{min}_\text{det}(\Lambda_p.\Gamma)$ depends on the number of continuous places, guard arcs and immediate and deterministic transitions, which are present in the model\add{, as discussed in Section~\ref{formalism}.}

%% file: sections/transient_domains.tex
\section{Transient analysis using interval arithmetic}
\label{ss:TA}
Transient analysis computes the probability to be in a certain state of the model at time $t'$. This can easily be extended to computing the probability that the model fulfills a certain atomic property, w.r.t. the discrete or continuous marking.  As the PLT is a symbolic representation of the state space evolution over time, the system can be in different parametric locations at time $t'$, depending on the values of the random variables. For each of these so-called \emph{candidate locations}, the subset of the potential domains need to be computed for which the system actually is in that location at time $t'$. The transient probabilities can then be computed by integrating over the computed subsets.
 
\add{This section is further organized as follows: 
Section~\ref{ss:candidates} explains how the candidate locations are obtained and their corresponding support is restricted. 
In Section~\ref{ss:transientprobs} the computation of the transient probabilities is detailed and Section~\ref{ss:numerical} discusses the numerical integration. The computational complexity of this approach is briefly discussed in the respective sections.}

 


\subsection{Candidate Locations and restricting their potential domains}
\label{ss:candidates}

Since the entry time $\Lambda.t$ of a location $\Lambda$ depends on the potential domain of the expired random variables $\Lambda.\mathbf{S}$, the entry time point is a linear function of the random variables.
However,  we can specify  the minimum entry time, i.e., the earliest possible entry time $\Lambda.t_\text{min}^\text{entry}$ of a  location $\Lambda$.
Accordingly, $\Lambda.t_\text{max}^\text{entry}$ specifies the latest possible entry time into a specific location. 
To identify the candidate locations for $t'$,  also the latest possible exit time of a location needs to be considered, which is obtained as the latest maximum entry time of all child locations:
\begin{equation}
    \Lambda.t_\text{max}^\text{exit} = \max\left\{\Lambda_c.t_\text{max}^\text{entry} \ \middle|\  \Lambda_c \in \text{children}(\Lambda) \right\}.
\end{equation}

 Note that, the number of locations in the tree for a maximum considered time $\tau_\text{max}$ is finite, according to \cite{hypro_valuetools}. 
To obtain candidate locations for a specific time  $t'$ we check for each location of the PLT, whether the potential domains are such that values for random variables exist, that bring the system in that location at time $t'$. 
 This is done by checking whether $\Lambda.t^\text{entry}_\text{min} \leq t'$ and $\Lambda.t_\text{max}^\text{exit} \geq t'$. If the first condition is violated, the entry time of the location $\Lambda$ is guaranteed to be later than the considered time point, and hence its children do \remove{also }not need to be considered as candidates\add{, either}. The second condition ensures that it is possible to still be in location $\Lambda$ at time $t'$. In  case this condition is violated, the child locations of $\Lambda$ still need to be considered  as candidates. Hence, the set of candidate locations $\mathcal{C}_{t'}$ at time $t'$ equals 
 \begin{equation} \mathcal{C}_{t'} = \left\{ \Lambda \in \mathbf{V} \ \middle|\ \Lambda.t_\text{min}^\text{entry} \leq t'\leq \Lambda.t_\text{max}^\text{exit}\right\}. \end{equation}
 
The earliest and latest possible entry time of a parametric location $\Lambda$ are computed iteratively by replacing random variables by their lower or upper boundaries. The boundaries of a random variable may depend on  random variables with a lower order, i.e., that fired earlier.  These dependencies need to be resolved when minimizing the entry time of a location or maximizing the entry time of its children. Using the previously defined order, the linear equation defining $\Lambda.t$ can be rearranged such that the expired random variables together with their factors occur according to their firing order: 
\begin{equation}\Lambda.t=f_0+\sum^m_{k=1}f_k \cdot \mathbf{o}[k].\end{equation}

By iterating over those random variables it is possible to resolve their dependencies and to minimize the entry time at the same time.
In case random variable $s_i^j$ is stored in $\mathbf{o}[l]$, the random variable $s_i^j$  needs to be replaced within $\Lambda.t$ by the corresponding interval bound $S_i^j.l$ or $S_i^j.u$, depending on the sign of $f_l$. This step is repeated for all $l \leq m$ in descending order and the resulting $\Lambda.t$ after each step then only depends on all $\mathbf{o}[k] \prec \mathbf{o}[l]$: 
\begin{align}
    \Lambda.t = 
    f_0 +\sum^{l-1}_{k=1}f_k \cdot \mathbf{o}[k] + \begin{cases} f_l \cdot S_i^j.l , & \quad \text{ if } f_l>0, \\
   f_l \cdot S_i^j.u , & \quad\text{ if }  f_l<0.
    \end{cases}
\end{align}

Once all random variables have been resolved by the corresponding lower or upper boundary, $\Lambda.t$ has been minimized within the potential domain of location $\Lambda$. When computing the maximum of the entry time for all children, i.e., the latest exit time point, the conditions to replace the random variables are both reversed. 





In all candidate locations the potential domain of all random variables  present up to $t'$ need to be restricted, such that only those values remain for which the system certainly is in that location at time $t'$. The computation of the restricted potential domains per location corresponds to a \emph{cylindrical algebraic decomposition} \cite{collins1991partial}, which returns  the restricted potential domains $\mathbf{S'}$ as a set of multi-dimensional intervals. For linear inequalities this can be computed in $\mathcal{O}\bigl(n^2\bigr)$ using a variant of the \emph{Fourier-Motzkin elimination} \cite{george2003fourier}.

Under the assumption that no conflicts have occurred, the restricted potential domain of a location and the restricted potential domains of its children form disjoint subsets. In case a conflict occurs, it needs to be  resolved by adapting the value $\Lambda_c.p$, for all children that participate in the conflict. In that case the restricted domains of all conflicting child locations may overlap. Note that, conflicting child locations have the same minimum entry time.

\subsection{Computing transient probabilities}
\label{ss:transientprobs}

When the candidate nodes and the restricted intervals at time $t'$ have been determined, the transient probabilities can be computed. Recall that each general transition was assigned a  continuous probability distribution which describes how the probabilities are distributed over the values of the random variables.
The probability to be in a specific candidate location $\Lambda$ at time $t'$ is computed by first integrating over the joint probability density function of all random variables, and then multiplying the result with the \emph{accumulated conflict probability} $ p_\text{acc}(\Lambda)$ towards that  location $\Lambda$. The latter is computed recursively by multiplying  the conflict probability of each location visited when traversing the tree from the root location to $\Lambda$:

\begin{align}
    p_\text{acc}(\Lambda) = 
     \begin{cases} 
    p_\text{acc}\left(\Lambda_p\right) \cdot \Lambda.p , & \quad\text{ if }\exists \Lambda_p: \Lambda \in \text{children}(\Lambda_p) , \\
     \Lambda.p , & \quad \text{$\Lambda$ is the root location}.
    \end{cases}
\end{align}

Recall that the potential domain $\Lambda.\mathbf{S}$ consists of one multi-dimensional interval for each location $\Lambda$. 
The restricted potential domain $\Lambda.\mathbf{S'}$ w.r.t. the time $t'$, however, may consist of several multi-dimensional intervals, as the Fourier-Motzkin elimination might split the potential domain in several dimensions. 
Hence, the restricted domain $\Lambda.\mathbf{S}'$ is a set of multi-dimensional intervals. 
Each element $\mathbf{r} \in \Lambda.\mathbf{S'}$ is represented by a vector of intervals defining an upper and a lower bound for each random variable present.
Note that, this does not influence the number of random variables present in location $\Lambda$: In case that $\left|\mathbf{s}_\Lambda\right| = n$ (expired and not yet expired) random variables are present in $\Lambda$, we obtain $n$-dimensional integrals. 
The random variables which correspond to the particular next firing of each enabled general transition need to be included to ensure that this firing has not taken place yet.

Recall that, having $m\leq n$ expired random variables, for each $k \in [1,m]$ and $\mathbf{r} \in \Lambda.\mathbf{S'}$, $\mathbf{r}[k]$ can only depend on the random variables $\mathbf{r}[1], \mathbf{r}[2], ..., \mathbf{r}[k-1]$. Similarly, for each $l \in [m+1, n]$, $\mathbf{r}[l]$ can only depend on $\mathbf{r}[1], \mathbf{r}[2], ..., \mathbf{r}[m]$.
Hence, the lower bound of $\mathbf{r}[1]$ is a constant as no dependencies can exist. For all $i \in [1, n]$,  the upper bound of $\mathbf{r}[i]$ might be infinite as well.
Further recall that the bounds of the (restricted) potential domains of each random variable are  linear expressions. We denote the lower bound of $r[i]$ with $r[i].l$ and the upper bound with $r[i].u$, respectively.

Let function $g_i(s)$ denote the probability density function of the random variables $\mathbf{s}_\Lambda[i]$, which is uniquely determined by the probability distribution assigned to the corresponding general transition. All random variables are accessed in their firing order, as specified in $\mathbf{s}_\Lambda$. The resulting transient probability at $t'$ for parametric location $\Lambda$ then equals the probability that the values of all random variables in $\mathbf{s}_\Lambda$ lie within $\Lambda.\mathbf{S'}$, multiplied by its accumulated conflict probability, as given by Equation \ref{eq:prob}:
\begin{equation}
\label{eq:prob}
Prob(\Lambda, \mathbf{s}_\Lambda, \Lambda.\mathbf{S'}) = p_\text{acc}(\Lambda) \cdot \sum_{\mathbf{r} \in \Lambda.\mathbf{S'}} \int_{\mathbf{r}[1].l}^{\mathbf{r}[1].u} \dots \int_{\mathbf{r}[n].l}^{\mathbf{r}[n].u} G(\mathbf{s}_\Lambda) \ \ \operatorname{d\mathbf{s}_\Lambda[n]} \cdots \operatorname{d\mathbf{s}_\Lambda[1]},
\end{equation} 
where\remove{as} $G(\mathbf{s}_\Lambda) =\prod_{i=1}^n g_i(\mathbf{s}_\Lambda[i])$ is the joint probability density function, which equals the product over the $n$ probability density functions, due to the independence of the random variables \add{as explained in Section~\ref{formalism}. 
Note that, multi-dimensional integration over dependent intervals is required, as the upper and lower bounds per random variable may depend on other random variables, as explained in Section~\ref{ss:expiredrv}.}


 \subsection{Transformation of variables} 
 \label{ss:transformation}

 There exist a variety of implementations for (numerical) multi-dimensional integration. However, to the best of our knowledge there are no libraries available for multi-dimensional integration over dependent intervals, but mostly  over (bounded) rectangular regions of integration, i.e., with constant limits only. To use e.g., the \textit{GNU Scientific Library (GSL)}\footnote{\url{https://www.gnu.org/software/gsl/}}~
 \cite{galassi}, which provides integration over rectangular regions, a transformation of each $\mathbf{r} \in \Lambda.\mathbf{S'}$ onto a rectangular region is required. Based on the approach of \cite{mcnamee}, we present such a transformation for the integral from Equation \ref{eq:prob} in the following.
 
 The basic idea is to use the Transformation Theorem \add{by Hammer and Wymore}~\cite{hammer1957NumericalEvaluationMultiple, bauer2011measure}, which is the generalization of integration by substitution on functions of higher dimensions.
 It is used for calculating integrals if the integral can be calculated more easily after transformation to another coordinate system. 
 The goal in the following is to use a transformation $\mathcal{T}$ to transform the arbitrary region of integration $V$ into a simpler, aligned polytope $V^*$ and to calculate the integral with integrand $f(x)$ by using the Transformation Theorem: 

\begin{equation}
\label{eq:transformation_theorem}
\int_{V} f(x) \operatorname{dx}
=
\int_{V^*} f(\mathcal{T}(\upsilon)) \cdot \left|{\det(D\mathcal{T}(\upsilon))}\right| \operatorname{d\upsilon},
\end{equation}
where $\mathcal{T}$ is a bijective transformation from $V^*$ to $V$ and $D\mathcal{T}$ is the Jacobian matrix of $\mathcal{T}$.
For Equation \ref{eq:transformation_theorem} to be applicable $f$ must be measurable on $V$ and $\mathcal{T}$ must be a diffeomorphism.
\replace{The f}{F}unction $f$ is measurable, since \replace{functions that are}{the} (product\remove{s} of) probability density functions over $V$ \replace{are}{is} measurable \remove{by definition}.
The transformation $\mathcal{T}$ is a differentiable, bijective function whose inverse is differentiable as well and hence fulfills the second condition.

Let $U^n$ be the $n$-cube with coordinates in $[-1, 1]$ for every dimension. A transformation $\mathcal{T}_\mathbf{r}$ for any $\mathbf{r} \in \Lambda.\mathbf{S'}$ with $|\mathbf{r}|=n$ from $U^n$ onto $\mathbf{r}$ is given by:

 

 \begin{equation}
\label{eq:transformation}
\begin{split}
\mathcal{T}_\mathbf{r} \colon U^n & \rightarrow \R^n, \\
\upsilon_i & \mapsto \frac{\mathbf{r}[i].u + \mathbf{r}[i].l}{2} + \upsilon_i \frac{\mathbf{r}[i].u - \mathbf{r}[i].l}{2} , \ \ \ i=1,\dots , n.
\end{split}
\end{equation}

 where\remove{as} $\upsilon_i$ denotes the variable transformed onto the random variable $\mathbf{s}_\Lambda[i]$. 
 
 Note that, this does not work for the case $\mathbf{r}[i].u = \infty$, as discussed later. 
 The Jacobian determinant of this transformation is given by:
 \begin{equation}
 \label{eq:jacobian}
 \left|{\det(D\mathcal{T}_\mathbf{r}(\upsilon))}\right| = \prod_{i=1}^n \left( \frac{\mathbf{r}[i].u - \mathbf{r}[i].l}{2} \right).
 \end{equation} 
 Hence, it follows:

 \begin{eqnarray}
Prob(\Lambda,\mathbf{s}_\Lambda, \Lambda.\mathbf{S'}) 
&=& p_\text{acc}(\Lambda) \cdot\sum_{\mathbf{r} \in \Lambda.\mathbf{S'}} \int_{\mathbf{r}[1].l}^{\mathbf{r}[1].u} \dots \int_{\mathbf{r}[n].l}^{\mathbf{r}[n].u} G(\mathbf{s}_\Lambda) \ \ \operatorname{d\mathbf{s}_\Lambda[n]} \cdots \operatorname{d\mathbf{s}_\Lambda[1]},\nonumber \\
 \label{eq:multiintegral}
 &=& p_\text{acc}(\Lambda) \cdot\sum_{\mathbf{r} \in \Lambda.\mathbf{S'}}  \int_{-1}^{1} \int_{-1}^{1} \cdots \int_{-1}^{1} \left(F(\boldsymbol{\upsilon}) \cdot \left|{\det(D\mathcal{T}_\mathbf{r}(\upsilon))}\right| \right) \ \operatorname{d\upsilon_n} \cdots \ \operatorname{d\upsilon_1}, \ \ 
 \end{eqnarray} 
 where $F(\boldsymbol{\upsilon})$ is $G(\mathbf{s}_\Lambda)$ under the transformation stated by Equation \ref{eq:transformation}.

Since the transformation of variables in \cite{mcnamee} is not defined for a random variable $s_i$ with an upper interval bound $\mathbf{r}[i].u = \infty$, we handle this special case as follows:
We make use of the maximum time $\tau_\text{max}$, up to which the PLT is created and in a first step we set $\mathbf{r}[i].u=\tau_\text{max}$. However, in case that the probability density of $s_i$ is greater than zero in the interval $[\tau_\text{max}, \infty]$, we further need to adjust the result of the integration to include the missing probability mass in $[\tau_\text{max}, \infty]$. This can be realized by both: (i) adding the integral over $s_i \in [0, \infty]$ and (ii) subtracting  the integral over $s_i \in [0, \tau_\text{max}]$, where the integrand and the interval bounds for all other bounded random variables remain the same.

\subsection{Numerical integration based on Monte Carlo}

\label{ss:numerical}
\remove{Numerical integration is often the only solution for the integration of unknown antiderivatives in multiple dimensions.}
\add{Solving Equation~\ref{eq:multiintegral} for unknown antiderivatives efficiently in multiple dimension requires numerical integration.} We use approximative methods together with an estimation of the (statistical) error for bounded rectangular regions. 

 Monte Carlo methods \cite{lepage} provide algorithms for the approximation of such integrals including a reliable error estimate, which are well-suited to compute integrals of higher dimensions since their convergence rates\add{, i.e., the speed at which the estimated probability approaches the real probability,} are independent of the dimension, c.f. \cite{lepage}. 
 The main idea for estimating an integral as in \cite{galassi,press} is to select $N$ random points from a distribution of points in the (rectangular) integration region $U^n$.
 
Let $x_i$ with $i \in \mathbb{N} : i \leq N$ denote those random points and let $V_{U^n}$ denote the volume of region $U^n$. An estimate $E(f;N)$ for the integral of the function $f(x)$ over the $U^n$ is then given by: 
\begin{equation}
\label{eq:estimate}
E(f;N) = \frac{V_{U^n}}{N} \sum_{i=1}^N f(x_i).
\end{equation}
The error on the estimate $\sigma(E;N)$ results from the square root of the estimated variance of the mean $\sigma^2(E;N)$, as stated by Equation \ref{eq:variance}:
\begin{equation}
\label{eq:variance}
\sigma^2(E;N) = \frac{(V_{U^n})^2}{N^2} \sum_{i=1}^N \left(f(x_i) - \langle f\rangle \right)^2,
\end{equation}
where $\langle f\rangle$ denotes the arithmetic mean of the evaluations of $f$ over $N$ sample points.

Our implementation uses the adaptive Monte Carlo scheme VEGAS \cite{lepage} as implemented in GSL to compute the transient probabilities. VEGAS is  based on \textit{importance sampling}, where the probability distribution of the random points is iteratively improved to reduce the variance.
We assume that a data structure $d$ is provided by implementation of the approach from Section~\ref{ss:candidates}. For every candidate node and for every firing in the past, i.e., for every random variable $s_i$, the corresponding functions $\mathbf{r}[i].l$ and $\mathbf{r}[i].u$ and its probability distribution are stored and subsequently transformed  onto the cube $[-1, 1]^n$ (\add{see}\remove{c.f.} Section~\ref{ss:transformation}). 

\add{Since the transformation of variables is done for every sample point, the total complexity of the transformation is in $\mathcal{O}(N\cdot n$).}
\add{When performing transient analysis, integration is performed once for each candidate location with a non-empty restricted potential domain $\mathbf{S}'$. When computing the probability for more complex properties, the restricted potential domain may consist of more intervals, whose number is bounded by the number of constraints considered in the cylindrical decomposition, which is given by $\mathcal{O}(2\cdot4^{n-2})$ for a location with $n$  random variables present~\cite{huls2020ModelCheckingHybrid}.}

%% file: sections/transient_geometrical.tex
\section{Geometrical approach to transient analysis} \label{ss:TGeo}
In addition to the state-space representation as PLT, a geometric representation of locations has been proposed together with an automated transformation from locations into geometric regions \cite{hypro_valuetools}. 
 The geometric state set representation relies on closed convex polytopes and the region corresponding to the location is constructed in $\mathcal{H}$-representation. This representation allows for model checking more complex properties, as described in \cite{huls2019ModelCheckingHPnGs}.

\add{In the following we extend the concept presented in Section~\ref{ss:TA} to the geometric representation of locations to facilitate the computation of probabilities
of more complex properties as required for model checking. Section~\ref{sec:regions} presents the concept of regions which can be represented using convex polytopes. Sections~\ref{ss:integconvex} and \ref{sec:numerical_integration_polytope} discuss two approaches on how to integrate over these polytopes to compute probabilities. The computational complexity of both is examined in Section~\ref{sec:std_complexity}. }


\subsection{The concept of regions}
\label{sec:regions}

The geometric representation of a location $\Lambda$ has $n+1$ dimensions, given a vector  $\mathbf{s}_{\Lambda}$ of size $n$. 
One additional dimension is included to represent  time. Following \cite{ghasemieh2012region}, a region is defined as the maximal connected set of states, whose discrete parts remain unchanged as long as no event occurs.

\begin{definition} A region $R$ is a maximal convex set of points $(\mathbf{s},t)$
, for which the following three conditions are fulfilled for all $(\mathbf{s_1},t_1),(\mathbf{s_2},t_2) \in R, (\mathbf{s_1},t_1)\not =(\mathbf{s_2},t_2)$:
\begin{enumerate}
    \item $\Gamma(\mathbf{s_1},t_1).\mathbf{m} = \Gamma(\mathbf{s_2},t_2).\mathbf{m}$,
    \item $\Gamma(\mathbf{s_1},t_1).\mathbf{d} = \Gamma(\mathbf{s_2},t_2).\mathbf{d}$,
    \item $\Gamma(\mathbf{s_1},t_1).\mathbf{e} = \Gamma(\mathbf{s_2},t_2).\mathbf{e}$.
\end{enumerate}
\end{definition}
Hence, within a region $R$, the discrete marking $\Gamma(\mathbf{s},t).\mathbf{m}$, the drift vector $\Gamma(\mathbf{s},t).\mathbf{d}$ and the enabling status vector $\Gamma(\mathbf{s},t).\mathbf{e}$ are equal for any point $ (\mathbf{s},t) \in R$. The amount of fluid and the clock valuations are linear equations of $\mathbf{s}$ and $t$. As a consequence, each boundary between regions, which represents the occurrence of an event, is characterized by a linear polynomial of $\mathbf{s}$ and $t$ and thus restricts a \emph{half-space} in $n+1$ dimensions. Due to the linearity of events and the restriction with respect to $\tau_\text{max}$, these half-spaces form convex polytopes. For a proof, we refer to   \cite{ghasemieh2017analysis}.
A polytope $P$ can be represented as the intersection of a finite set of half-spaces stored as a pair of a matrix $A$ and a vector $b$ such that $P = \{ x \mid A\cdot x \leq b \}$. This representation is referred to as $\mathcal{H}$-representation and the resulting polytope is called $\mathcal{H}$-polytope. 

 \paragraph{Transformation into regions and implementation}
 
A transformation from locations into geometric regions has been presented in \cite{hypro_valuetools} which fully relies on  halfspace intersection, which is dual to the convex-hull problem.  We used the C++ library \hypro~\cite{schupp2017hypro} to create a convex polytope representing a region using \replace{quick hull}{quickhull} \cite{barber1996quickhull} algorithms. This allows the analysis of Hybrid Petri nets with a finite number of multiple stochastic firings. 
\hypro~does not limit the number of dimensions and provides most of the set operations required for the analysis of HPnGs. We are aware that there exist various implementations of convex polytopes, e.g., \ppl~\cite{BagnaraHZ08SCP}, \polymake~\cite{gawrilow2000polymake}, \cgal~\cite{cgal:eb-17a}, \cdd~\cite{cddWebsite}, which can be used as well but do not provide intuitive, convenient interfaces and conversion functions. Furthermore we can exploit the features of \hypro~for model checking.




Computing the transient probability for a given time point $t'$ requires intersecting every region with the  hyperplane representing time $t'$.  As convex polytopes are closed w.r.t. intersection, the result (if not empty) is again a convex polytope  that represents the possible states of the system at time $t'$. \add{To compute the joint probability distribution the dimension corresponding to time is removed from the polytope representation by projection.}

\subsection{Integration over convex polytopes}\label{ss:integconvex}
The integration bounds for polytopes that are aligned to the coordinate axes can be trivially deduced from the vertices. In general, however, polytopes created for a parametric location are not aligned. To integrate over general polytopes, two different approaches are discussed in this Section. The first approach is based on integration over general simplices as presented in Section \ref{sec:algorithm_base_case}.  
To use this integration method, Section \ref{sec:algorithm_partitioning} presents how an arbitrary convex polytope can be divided into standard simplices. The presented approach maintains a strict order of the integration intervals and allows the use of general integration methods. Alternatively,  Section \ref{sec:numerical_integration_polytope} shows how Monte Carlo integration can be applied directly to convex polytope using a bounding box. 

\subsubsection{Transformation to simplices}

\label{sec:algorithm_base_case}

Computing the Jacobian Matrix $D\mathcal{T}$ as presented in Section~\ref{ss:transformation} for an arbitrary transformation $\mathcal{T} \colon \R^n \rightarrow \R^n$ is computationally expensive, due to the quadratic nature of the matrix. Alternatively, an affine transformation may be used \cite{bauer2011measure}. An affine transformation $A\cdot R +b$ for a $d$-dimensional set $R$ with $A \in \R^{d \times d}$ and $b \in \R^d$ is defined as the set:

$$\{A \cdot x + b \: \vert \: x \in R\},$$

and can be described as a combination of scalings, rotations and skewings by a matrix $A \in \R^{d \times d}$ and a translation given by the vector $b \in \R^d$. 
For such an affine transformation $\mathcal{A}$ the Jacobian matrix $D\mathcal{A}$ simply is $A$, hence, only   the determinant of a matrix  needs to be calculated. 
Additionally, $\mathcal{A}$ is a diffeomorphism if $A$ is invertible because all linear functions $A\cdot x$ are differentiable. 
However, a limiting factor of affine transformations is their preservation of parallelism, which greatly limits the types of polytopes that they can align to the coordinate axes. 
For example, a general polytope with four vertices cannot be affinely transformed into a square because the opposing sites of a quadrilateral are, in general, not parallel. 
A polytope for which such an affine transformation is always possible is a so-called $n$-simplex. 
A simplex is the generalization of the notion of a triangle to arbitrary dimensions. 
A unit $n$-simplex $\Delta_n^* \subseteq \R^n$ is defined by the convex hull $cHull(\{\mathbf{0},e_1,\dots,e_n\})$ of its $n+1$ vertices denoted $\{\mathbf{0},e_1,...,e_n\}$:
    
\begin{equation}
\label{eq:unit_simplex}
\begin{split}
	\Delta_n^* \coloneqq& \ \conv\left(\left\{\mathbf{0}, e_1, \dotsc , e_n \right\}\right) \\
	=& \ \left\{ x \in \R^n \;\middle|\;\sum_{i=1}^n x_i \leq 1 \land \forall i \in \{1,\dots,n\} \colon x_i \geq 0 \right\}.
\end{split}
\end{equation} 
For a unit $n$-simplex $\Delta_n^*$  the region of integration is given by the set of intervals as presented in the following.  The integration limits are either constant or depend on previous variables, according to Section \ref{ss:transientprobs}. This order is given for unit simplices, as shown by the following equation:

\begin{equation}
\label{eq:unit_simplex_integration_intervals}
\forall x \in \Delta_n^* \colon \forall i \in \{1,\dots,n\} \colon x_i \in \left[0, 1-\sum_{j=1}^{i-1}x_j\right].
\end{equation}
Each coordinate is non-negative and must be at most one minus all of the preceding coordinates accumulated.
This ensures that the total sum of the coordinates is at most one and that the sequence of intervals describes exactly the set given in Equation \ref{eq:unit_simplex}.

Finally, for any n-simplex $\Delta_n = cHull(\{v_0, \dots, v_n\}) \subseteq \mathcal{R}^n$, an affine transformation $\mathcal{A}_{\Delta_n}$ from the unit n-simplex $\Delta_n^*$ onto $\Delta_n$, is defined by
\begin{equation}
\label{eq:simplex_trans}
\begin{split}
\mathcal{A}_{\Delta_n} \colon \R^n & \rightarrow \R^n \\
x & \mapsto \left((v_1 - v_0)\dots(v_n - v_0)\right) \cdot x + v_0.
\end{split}
\end{equation}

The exact assignment of a random variable $s_{i}^{j}$ in a vertex $x$ is denoted $s_{i}^{j} \in x$. Recall that the function $g(s_{i}^{j})$ denotes the probability density function of a random variables $s_i^j$, as presented above. The joint probability density function of all random variables present in a parametric location is denoted as $G(x)$ in the following. Since the firing times of general transitions are independently distributed,  the joint probability of all random variables represented by a vertex $x$ is the product of their individual densities, i.e.,

\begin{align}
   G(x) = \prod_{\forall s_{i}^{j} \in x} g(s_{i}^{j}).
\end{align}

For any simplex $\Delta_n \in \satt[t](\varPhi)$ the probability over the joint probability density function $G$\ can be computed as:
\begin{gather}\label{eq:integral_over_transformed_unit_simplex}
\int_{\Delta_n} G(x) \operatorname{dx} =
\int_{0}^{1}\dots\int_{0}^{1 - \sum_{i=1}^{n - 1}x_i}
G(\mathcal{A}_{\Delta_n}((x_1,\dots,x_n))) \cdot \left|{\det(A)}\right| \operatorname{dx_n}\dots\operatorname{dx_1},
\end{gather}
with normalized integration bounds.
The affine transformation $\mathcal{A}_{\Delta_n}$  is defined by the  $n \times n$ matrix  $A$,  as defined in Equation \ref{eq:simplex_trans}.

\begin{algorithm}[tb]
	\ttfamily 
	\begin{algorithmic}[1]
		\caption{getProbability(s: $\triangle$)}
        \label{alg:prob_simplex}
        \STATE matrix $\add{A}$ : $\mathbb{R}^{n \times n}$
        \STATE $\{v_0,...,v_n\}$ = s.getVertices()
        \FORALL{column $A_i$ of $A$}
            \STATE $A_i$ = $(v_i-v_0)$
        \ENDFOR
        \STATE $\mathcal{A} = \mathbf{s} \rightarrow A\cdot \mathbf{s} + v_0$
        \STATE $G = \mathbf{s} \rightarrow \prod_{i=1}^n g(\mathbf{s})$
        \STATE $f = \mathbf{s} \rightarrow G(\mathcal{A}(\mathbf{s})) \cdot \vert \text{det}(A) \vert$
        \STATE $I$ : list()
        \FOR{$i=0$ \textbf{ to } $\mathbf{o}$.size()}
            \STATE $I$.append($[0,1-\sum_{j=0}^i \mathbf{o}[j]]$)
        \ENDFOR 
        \STATE return integrate($f$,$I$)
	\end{algorithmic}
\end{algorithm}


Algorithm \ref{alg:prob_simplex} provides pseudo-code for the function \texttt{getProbability} to obtain the probability to be in a simplex, with a single function parameter s. First, the matrix $A$ needed for the affine transformation is initialized (line 1) and all vertices are obtained from the simplex s (line 2). These vertices are used to create the transformation matrix (line 3-4) as presented in Equation \ref{eq:simplex_trans}. Then three functions are initialized, each of which return values for a vector of random variables. The function $\mathcal{A}$ denotes the affine transformation (line 5), the function $G$ returns the joint probability density of all random variables (line 6) and the function $f$ is used to apply the transformation theorem during integration (line 7). The region of integration then is initialized (line 8-10) as presented in Equation \ref{eq:unit_simplex_integration_intervals}. To obtain the probability, a function \texttt{integrate} is called with the variables $f$ and $I$. Hence, the integration method can easily be exchanged. 

\subsubsection{Extension to arbitrary convex polytopes}
\label{sec:algorithm_partitioning}
Algorithm~\ref{alg:prob_simplex}  computes $\prob(\Delta_n)$ for arbitrary simplices $\Delta_n$ and needs to be  generalized to arbitrary convex polytopes $R$. To compute the probability $\prob(\Region)$, the polytope $\Region$ needs to be subdivided into simplices for which the probability can be calculated individually. In the following a polytope is partitioned using triangulation denoted as $\triangulation(\Region)$. More formally, the $n$-dimensional Delaunay triangulation \cite{toth2017handbook} is used as presented in the following.

\begin{definition}[n-dimensional Delaunay-triangulation]
    Given a convex polytope $\Region \subseteq \R^n$ in $n$-dimensional space, the triangulation $\triangulation(\Region)$ is a simplicial complex 
    \begin{align*}
        \triangulation(\Region) := \{\Delta \subseteq \R^n\: \vert \: \Delta \text{ is a simplex}\}\add{,}
    \end{align*}
    \add{where $\Region$ is the union of all simplices, i.e., $\Region = \bigcup_{\forall \Delta \in  \triangulation(\Region)} \Delta$, }and the intersection of two simplices is either empty or a facet, i.e., $\forall \Delta_i, \Delta_j\in \triangulation(\Region)$ \add{ with $i \neq j$}, $\Delta_i$ and $\Delta_j$ are disjoint. 
\end{definition}

The probability to be in $\Region$ can then be computed in three consecutive steps:

\begin{enumerate}
	\item Construct a triangulation $\triangulation(\Region)$ for $\Region$.

	\item For each resulting simplex  $\Delta \in \triangulation(\Region)$, calculate the probability to be in that simplex, i.e., $\prob(\Delta)$, using Algorithm \ref{alg:prob_simplex}.

    \item The total probability can be accumulated, as the resulting simplices  are pairwise disjoint:

      $\prob(\Region) = \sum_{\Delta_i \in \triangulation(\Region)} \prob(\Delta_i).$
     
\end{enumerate}

\subsection{Direct numerical integration over convex polytopes} 
\label{sec:numerical_integration_polytope}
The previously presented effort to generate integration regions with very specific boundaries  from arbitrary convex polytopes allows the use of different  integration methods. 
When focusing on numerical integration methods, the integration can be carried out  directly on convex polytopes, as  presented in the following. This concept is implemented via Monte Carlo integration as included in the GSL library and estimates  the multidimensional integral of the form 
\begin{align*}
    \int_{x_1.l}^{x_1.u}  \dots  \int_{x_n.l}^{x_n.u} f(x_1,\dots, x_n) \operatorname{dx_n}\dots\operatorname{dx_1}
\end{align*} 
over a hypercubic region $((x_1.l,x_1.u), \dots, (x_n.l,x_n.u))$. The minimum axis-aligned bounding box of a convex polytope can be used as a hypercubic region for Monte Carlo integration, as defined below.

\begin{definition}[Axis-aligned bounding box]
    Let $P \subseteq \R^n$ be a convex polytope. The bounding box of a convex polytope $P$ is an axis-aligned box within which all points of $P$ lie. The minimum value in a dimension $i$ is obtained by $\min_{x \in P} x_i$ and the maximum value in a dimension $i$ is obtained by $\max_{x \in P} x_i$. 
    The bounding box $\operatorname{bb}(P)$ of a polytope $P$ then is defined as the cartesian product of $n$ intervals:
    \begin{align*}
        \operatorname{bb}(P) := [\min_{x \in P} x_1, \max_{x \in P} x_1] \times \dots \times [\min_{x \in P} x_n, \max_{x \in P} x_n].
    \end{align*} 
\end{definition}

The bounding box of a candidate region $\Region$ defines the region of integration.   To compute the probability over a candidate region $\Region$, an indicator function evaluates whether a point $x$ lies within $\Region$:
\begin{align*}
    \mathbf{1}_{\Region}(x) := \left\{
\begin{array}{ll}
1, & \text{if } x \in P, \\
0 & \text{otherwise}. \\
\end{array}
\right.
\end{align*}
Using the joint probability density function $G(x)$ as presented above, the probability for a point $x$ to be in a candidate region $\Region$ can be computed as:

\begin{align*}
    \int_{\operatorname{bb}(\Region)} \mathbb{1}_{\Region}(x) \cdot \operatorname{G}(x) \operatorname{dx}.
\end{align*}

Hence, the Monte Carlo integration samples a vertex $x \in \operatorname{bb}(\Region)$ and the joint density $G(x)$ is only accumulated in case the vertex $x$ lies within $\Region$. 


\subsection{\add{Complexity}}
\label{sec:std_complexity}

\add{
In the following, we first discuss the complexity of transforming a single parametric location into a region, as briefly explained in Section~\ref{sec:regions}. Then we indicate the computational complexity of integration over convex polytopes, relying on triangulation, as indicated in Section~\ref{ss:integconvex}. Finally, we discuss the complexity when integrating directly over convex polytopes using a bounding box, as explained in Section~\ref{sec:numerical_integration_polytope}.
}

\add{
\paragraph{Constructing a region.} Since the construction of a region solely relies on the geometric intersection of half-spaces, we limit our discussion to the estimation of the number of intersections that is required.  As presented in \cite{ziegler1995LecturesPolytopes}, intersection of polytopes in $\mathcal{H}$-representation can be performed in polynomial time, but heavily depends on the number of dimensions.
A single intersection is performed for the source event of the location and multiple intersections for every child of the location. According to Section \ref{ss:childlocations}, the number of children of a parametric location $\Lambda$ is in }

\begin{equation}
\label{eq:complexity}
\add{\mathcal{O}\Bigl(\left\vert \mathcal{E}^\text{min}_\text{det}(\Lambda.\Gamma)  \right\vert +n\Bigr).} 
\end{equation}

\add{
Further, half-spaces are created for the constraints of the potential domain of a parametric location $\Lambda$. According to Section \ref{ss:childlocations}, the number of random events, i.e., general transition firings, that may occur is in $\mathcal{O}(n)$ with $n = \vert \Lambda.\mathbf{S} \vert$. This results in $\mathcal{O}(2\cdot n)$ half-spaces, i.e, upper and lower bounds for each potential domain. However, this does not change the worst case complexity, denoted in Equation~\ref{eq:complexity}. }


\add{
\paragraph{Pre-computations required for geometric integration.}
Geometrically, the restricted domains required for integration are computed by halfspace intersection followed by a projection to remove the dimension of time from the polytope representation. Internally, this requires to change the $\mathcal{H}$-representation to a $\mathcal{V}$-representation and back. Relying on \hypro, this is realized by the quickhull algorithm \cite{barber1996quickhull}. The worst case complexity of this algorithm is quadratic with a linear space complexity in the worst case.
}

\add{
\paragraph{Integration over simplices.}
For determining the complexity of the integration over simplices, we consider the complexity of the different steps. For a $d$-dimensional polytope with $n_f$ facets, the worst-case computational complexity of the Delauney triangulation is in $\mathcal{O}\left((n_f)^2\right)$ and results in $\mathcal{O}\left((n_f)^{\frac{d}{2}}\right)$ simplices~\cite{toth2017handbook, seidel1995UpperBoundTheorem}. Note that, the number of facets $n_f$ is bounded by the number of events, which may occur in a parametric location.
}

\add{
Since Monte-Carlo integration has to be performed for each simplex, the resulting number of integrations is in
} 
\begin{equation}
\add{
\mathcal{O}\Bigl(\left(\left\vert \mathcal{E}^\text{min}_\text{det}(\Lambda.\Gamma)  \right\vert+n\right)^{\frac{n}{2}}\Bigr).
}
\end{equation} 

\add{
\paragraph{Direct integration.}
 The direct integration of a region requires the computation of one bounding box for the set of convex polytopes per candidate location. This has an exponential worst case complexity in the number of dimensions.  Integration is then called once for the area defined by each bounding box, and a containment test is required for each integration run to indicate whether the respective sample lies in the currently considered set of convex polytopes.
Containment testing has a worst case complexity which is linear in the number of half-spaces.}

%% file: sections/casestudy.tex
\section{Case Study: Battery back-up}\label{ss:CST} 

\add{
This section presents a case study model of a factory with battery back-up in Section~\ref{ss:moddes} and validates the methods discussed in this paper with statistical model checking in Section~\ref{ss:compres}, before discussing their performance in more detail in Section~\ref{ss:performance}.  Section~\ref{sec:CPU-memory} presents additional measurements on CPU and memory usage and allows to identify the current bottleneck of the approach. }


\subsection{\add{Model description}}\label{ss:moddes}
The system model presented in the following is an HPnG model of a factory with battery back-up. We carry out transient analysis to evaluate the availability of power supply from the grid and to show the feasibility of the presented approach. The analysis is carried out end-to-end, i.e., the model description is parsed, the state space is created,  transient analysis is performed and the resulting probability is returned.

As a feasibility study, we model a factory whose power intake from the grid is limited by a service level agreement (SLA) with the energy provider. The factory relies on battery back-up which is discharged in the case of a peak-demand or a power outage. 
Figure \ref{fig:batterymodel} shows the core of the HPnG model\add{, consisting of the battery, the power grid and costs, and Figure~\ref{fig:demands} shows the power demands}.
The continuous place $P^c_0$ \add{in Figure~\ref{fig:batterymodel}} models the battery with capacity $B$, an inflow  ($T^\mathit{Dy}_0$) from the power grid and an outflow ($T^\mathit{Dy}_1$) which corresponds to the power that exceeds the constant peak demand $P_A$, specified by the SLA.  
The rates of the dynamic inflow and outflow  are specified as  
$f_{T^\mathit{Dy}_0}\left(\mathbf{d}\right) = \max\left(\left(P_A - \sum_{d_i \in \mathbf{d}} d_i\right),0\right) $
and  $f_{T^\mathit{Dy}_1}\left(\mathbf{d}\right) = \max\left(\left(\sum_{d_i \in \mathbf{d}} d_i - P_A\right),0\right)$, 
where $\mathbf{d} \in \mathbb{R}^{|\mathcal{T}^{F}|}$ is the vector of currently active demands \remove{, as specified in Figure 4} \add{(which are modeled in Figure~\ref{fig:demands})}. \remove{All arcs (except the guard and inhibitor arc connected to $P^c_0$) have weight one, which is not indicated in the Figures.}


The upper left part of Figure~\ref{fig:batterymodel} models the status of the grid \add{(indicated by a box)} and enables the inflow of the battery. \add{Initially, the grid is on, which is indicated by the token in $P^d_0$}. The failure of the grid ($T^G_0$) is modeled by a general transition\add{, moving the token to $P^d_1$}, \remove{the recovery $T_0^D$ is deterministic} \add{indicating that the grid is off}.  \add{$T_0^D$ models the deterministic recovery of the grid.}

\add{Initially, the place $P^d_2$ holds a token, which indicates that the battery still has power. When the battery is completely drained, the immediate transition $T^I_0$ fires and moves the token to place $P^d_3$, such that $T^I_1$ gets enabled. When the battery is filled again, $T^I_1$ moves the token back to $P^d_2$. Hence, a token in}
\remove{place} $P^d_3$ indicates an empty battery, which in turn enables \add{the dynamic} transition  $T^\mathit{Dy}_2$. This \add{transition} models the cost \add{flow} that occur due to service level violations when the accumulated demand exceeds the amount specified in the SLA.  Its flow rate is  $f_{T^\mathit{Dy}_2}\left(\mathbf{d}\right) = \max\left(\left(\sum_{d_i \in \mathbf{d}} d_i - P_A\right),0\right)$ and the extra cost is \add{then} collected in the continuous place $P^c_1$ (with infinite capacity). 

Figure \ref{fig:demands} models three levels of demand via  static continuous transitions, i.e., standard ($T^F_{d_{1}}$),  reduced ($T^F_{d_{0}}$), and  extended ($T^F_{d_{2}}$), that get enabled via \remove{a }guard arcs \add{which are connected to the discrete places $P^d_{d_0}$, $P^d_{d_1}$ and $P^d_{d_2}$}. General transitions \add{between those places} switch from standard to reduced demand ($T^G_{d_{10}}$) and from standard to extended demand ($T^G_{d_{12}}$). 
\add{Deterministic transitions model the possibility to switch back to standard demand (from reduced with $T^D_{d_{01}}$ and from extended demand with $T^D_{d_{21}}$).
The token indicates the current demand by enabling one of the corresponding continuous transitions.}

The number of random variables in the system depends on the considered time $t'$ of the transient analysis and the repair and switching times. At least three random variables are present, since all three general transitions are initially enabled concurrently. Hence\add{, the deterministic firing times of transition} $T_0^D$, \add{modeling the repair of the grid, and transitions} $T^D_{d_{01}}$ and $T^D_{d_{21}}$\add{, which reset the power demand to standard,}  have a high impact on the complexity of the model.


\begin{figure*}
        \input{images/batterymodel.tex}
        \caption{HPnG model of a factory that is connected to the power grid and has an additional battery storage.}
        \label{fig:batterymodel}
\end{figure*}
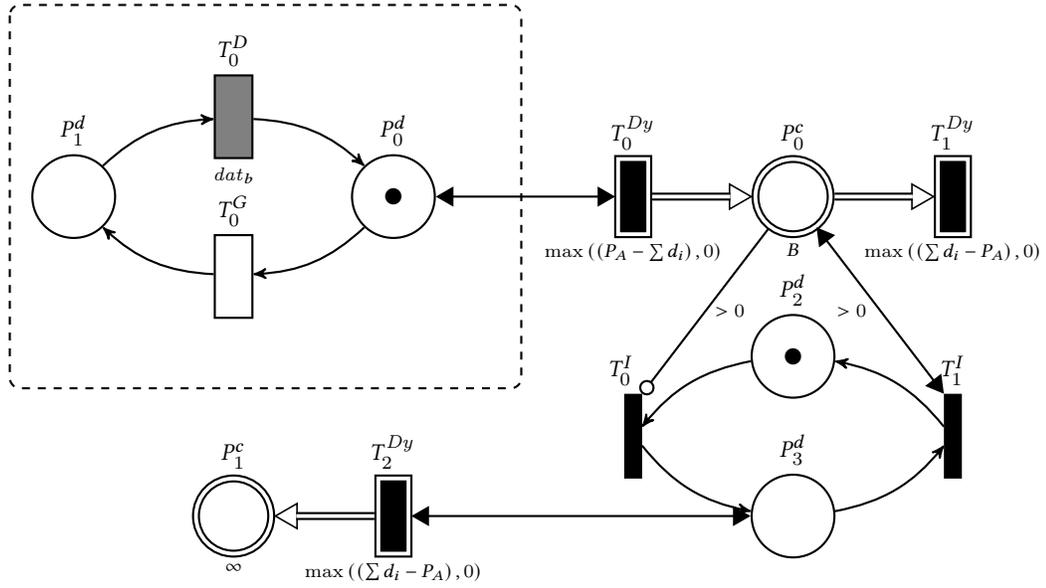
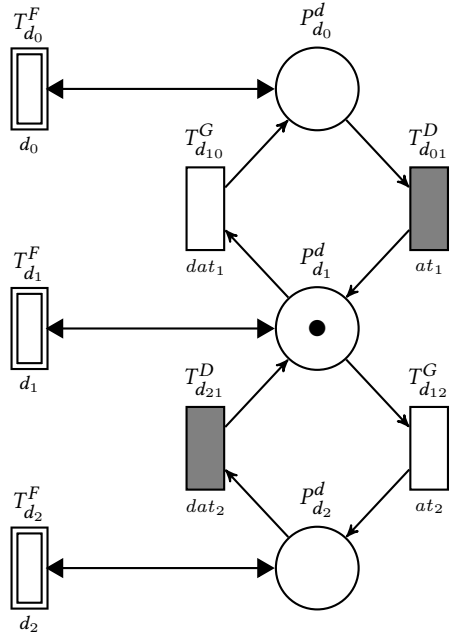
\begin{figure*}
        \input{images/demands.tex}
        \caption{HPnG model for the different power demands of the factory.}
        \label{fig:demands}
\end{figure*} 


 The parameters of the \replace{single}{specific} model components are presented in Table~\ref{tab:parameters}, as used in the initial setting of the case study. All of the general transitions follow a uniform distribution $\mathcal{U}\left(\SI{0}{\hour},\SI{10}{\hour}\right)$ and the firing time of $T_0^D$ takes natural values between $\SI{8}{\hour}$ and $\SI{1}{\hour}$.  \remove{resulting in an increasing availability of the grid. } \add{All arcs (except the guard and inhibitor arc connected to $P^c_0$) have weight one, which is not indicated in the figures.}

\begin{table}[tb]
    \centering
    \begin{tabular}{|p{5cm}|p{2cm}|}
     \hline 
     \textbf{Parameter} & \textbf{Value}\\
        \hline  \hline 
        Peak demand $P_A$ & $\SI{700}{\kW}$\\ \hline 
        Initial level $m_0(P^c_0)$ & $\SI{1000}{\kWh}$ \\ \hline 
        Capacity $B$ of $P^c_0$ & $\SI{1500}{\kWh}$\\ \hline 
        Initial level $m_0(P^c_1)$ & $\SI{0}{\kWh}$ \\ \hline 
        Continuous rate $d_0$ of $T^F_{d_0}$& $\SI{500}{\kW}$ \\ \hline 
        Continuous rate $d_1$ of $T^F_{d_1}$ & $\SI{700}{\kW}$ \\ \hline 
        Continuous rate $d_2$ of $T^F_{d_2}$ & $\SI{800}{\kW}$ \\ \hline 
        Distribution of $T^G_0$ & $\mathcal{U}(\SI{0}{\hour},\SI{10}{\hour})$ \\ \hline 
        Distribution of $T^G_{d_{10}}$ & $\mathcal{U}(\SI{0}{\hour},\SI{10}{\hour})$ \\ \hline 
        Distribution of $T^G_{d_{12}}$ & $\mathcal{U}(\SI{0}{\hour},\SI{10}{\hour})$ \\ \hline 
        Firing time $\text{at}_1$ of $T^D_{01}$ &$\SI{11}{\hour}$\\\hline
        Firing time $\text{dat}_2$ of $T^D_0$ & $\SI{11}{\hour}$\\\hline
    \end{tabular}
    \caption{Parameters used for the single components of the case study model in its initial setting. \add{The discrete initial markings are indicated in the figure.}}
    \label{tab:parameters}
\end{table}

\subsection{\add{Computation and validation of results}} \label{ss:compres}

The probability that certain properties hold is computed separately using all three methods previously presented, which were implemented within the tool \hpnmg~\cite{huels2020hpnmg}. The method \textbf{NUM} \textit{intervals} performs integration  on the intervals computed from the PLT, \textbf{NUM} \textit{simplices} integrates over general polytopes transformed to standard simplices. \textbf{NUM} \textit{polytopes} is a direct integration on convex polytopes. All three methods rely on Monte Carlo integration, for which error estimates are computed. For comparison, we include results obtained from the statistical model checker \hypeg \cite{pilch2017HYPEGStatisticalModel}, whose technique is based on simulation and thus denoted as \textbf{SIM} in the following.

\begin{table}[h]
    \centering
    \begin{tabularx}{\linewidth}{|l|p{1.5cm}|Y|Y|Y|Y|Y|Y|}
        \hline 
         $T_0^D$ & & $\SI{8}{\hour}$ & $\SI{7}{\hour}$ & $\SI{5}{\hour}$ &$\SI{3}{\hour}$ & $\SI{2}{\hour}$& $\SI{1}{\hour}$\\
         \hline
         \hline
         & dim & $4$ & $4$ & $5$ & $6$ & $8$ & $13$ \\ 
         \hline
         \multirow{2}{*}{\shortstack[l]{\textbf{NUM} \\ \textit{intervals}}}
         & $p\left(t',\Phi_1\right)$ & $0.200$ & $0.295$ & $0.454$ & $0.593$ & $0.689$ & $0.817$ \\
         & $e\left(t',\Phi_1\right)$ & 0 & $5.5\cdot e^{-6}$ & $5.8\cdot e^{-6}$ & $1.2\cdot e^{-5}$ & $1.0\cdot e^{-5}$ & $1.9\cdot e^{-5}$ \\ 
         & comp. time & $\SI{1.67}{\second}$ & $\SI{17.74}{\second}$ & $\SI{38.92}{\second}$ & $\SI{90.22}{\second}$ & $\SI{250.15}{\second}$ & $\SI{2145.66}{\second}$ \\
         \hline
         \multirow{2}{*}{\shortstack[l]{\textbf{NUM} \\ \textit{simplices}} }
         & $p\left(t',\Phi_1\right)$ & $0.200$ & $0.295$ & $0.455$ & $0.594$ & & \\
         & $e\left(t',\Phi_1\right)$ & $5.2\cdot e^{-6}$ & $7.7\cdot e^{-6}$ & $6.1\cdot e^{-5}$ & $3.9\cdot e^{-5}$ & - & - \\ 
         & comp. time & $\SI{2.85}{\second}$ & $\SI{5.66}{\second}$ & $\SI{33.45}{\second}$ & $\SI{344.54}{\second}$ &  &  \\
         \hline
         \multirow{2}{*}{\shortstack[l]{\textbf{NUM} \\ \textit{polytopes}} }
         & $p\left(t',\Phi_1\right)$ & $0.200$ & $0.295$ & $0.455$ & $0.593$ & $0.691$ &  \\
         & $e\left(t',\Phi_1\right)$ & $0$ & $4.2\cdot e^{-6}$ & $2.2\cdot e^{-4}$ & $5.6\cdot e^{-4}$ & $9.6\cdot e^{-4}$ & - \\ 
         & comp. time & $\SI{0.77}{\second}$ & $\SI{1.53}{\second}$ & $\SI{3.94}{\second}$ & $\SI{21.78}{\second}$ & $\SI{1721.95}{\second}$ & \\
         \hline
         \hline
         \multirow{2}{*}{\shortstack[l]{\textbf{SIM}}} & $p\left(t',\Phi_1\right)$ & $0.200$ & $0.296$ & $0.455$ & $0.594$ & $0.690$ & $0.818$ \\
         & comp. time \newline \tiny{(CI: $\pm 0.001$)}  & $\SI{27.61}{\second}$ & $\SI{31.30}{\second}$ & $\SI{39.87}{\second}$ & $\SI{39.29}{\second}$ & $\SI{33.57}{\second}$ & $\SI{26.04}{\second}$ \\
         & comp. time \newline \tiny{(CI: $\pm 0.0001$)}  & $\SI{1477.39}{\second}$ & $\SI{1983.97}{\second}$ & $\SI{2599.05}{\second}$ & $\SI{2761.17}{\second}$ & $\SI{2630.43}{\second}$ &  $\SI{1728.41}{\second}$ \\ 
         \hline
    \end{tabularx}
    \caption{Probability of an available grid, i.e., $\Phi_1:=m\bigl({P_0^d}\bigr) = 1$, for  $t'=\SI{8}{\hour}$ and decreasing grid repair times $T_0^D$, computed numerically with different integration approaches and via simulation. We state the computed probability $p(t',\Phi_1)$, the error from Monte Carlo integration $e(t', \Phi_1)$ and the computation time.}
    \label{tab:verify_simulation_time}
\end{table}

 
\add{In this subsection we compute the transient probability for two different measures, namely (i) the probability that the grid is available at time $t'=\SI{8}{\hour}$ for decreasing grid repair times and (ii) the probability that the standard demand is enabled at time $t'=\SI{8}{\hour}$ for randomly distributed demand changing times with different distributions. } 

 \add{Results for the first property obtained by the three numerical methods are summarized in   Table \ref{tab:verify_simulation_time} and validated with the statistical model checker \hypeg.}
  Each column \replace{of the}{in this} table represents \replace{one of these}{a different} value\remove{s} for the firing time of the deterministic transition\replace{.}{, which models the duration of grid repair.}  \add{Every entry in} Table \ref{tab:verify_simulation_time} \add{then} shows the probability that $P_0^d$ contains \replace{a}{one} token at time $t'=\SI{8}{\hour}$, \add{indicating that the (possibly repaired) power grid is available}.
 
 The number of random variables present in the system increases,  since $T_0^G$ can fire more often with decreasing firings times of transition $T^D_0$. The first row of the table represents the resulting state-space dimension. The other rows show the results for the three different numerical computation approaches and the simulation. A cell represents the results of the transient analysis verifying that $\Phi_1 := m\Bigl({P_0^d}\Bigr) = 1$ 
 holds at time $t'=\SI{8}{\hour}$, where $m\Bigl({P_0^d}\Bigr)$ denotes the marking of the discrete place $P_0^d$. 
 The resulting probability is denoted $p\left(t',\Phi_1\right)$ and the resulting error is denoted $e\left(t',\Phi_1\right)$. The last value in every cell is the computation time. Note that, simulation is executed for two $99\%$ confidence intervals with two different widths, i.e., $\pm 0.001$ and $\pm 0.0001$. 

The numerical results are well supported by the simulation \textbf{SIM} and illustrate that the computation times of simulation are independent of dimensions, whereas the numerical computation times grow exponentially with increasing dimension. While the \textbf{NUM} \textit{intervals} method is able to compute the probability for a model setting with up to $13$ dimensions,  the simplices method can not provide results for the model with more than $6$ dimensions and the polytope method can only deal with up to $8$ dimensions. In the cases indicated with -, the computation was terminated after two hours as the integration was taking too much time. 

\textbf{NUM} \textit{polytopes} performs better except for the two highest dimensions. For at most $6$ dimensions it outperforms the other methods by an order of magnitude. \textbf{NUM} \textit{simplices} takes longer than direct integration over polytopes due to the required transformation to standard simplices, which results in a larger number of objects which are each integrated over individually. The  advantage of the simplices method is however, that it would allow the use of other multi-dimensional integration methods. While the \textbf{NUM} \textit{intervals} method performs best in this setting,  it can  only be used to compute transient distributions. Model checking, as proposed in \cite{huls2019ModelCheckingHPnGs}  requires geometric operations and hence, integration on polytopes.

The number of dimensions affects the computation time of the numerical analysis, whereas the width of the confidence interval significantly influences the computation time of the simulation. A confidence interval of $\pm 0.00002$ would be comparable to the largest error made by the numerical analysis. This is however not feasible, since the simulation takes too long for this confidence interval width. For $\pm 0.001$ and $\pm 0.0001$, the statistical simulation takes (significantly) more time than the numerical analysis for four or five dimensions.  For 13 dimensions, the computation time of numerical analysis and simulation with a confidence interval of $\pm0.0001$ are  of the same magnitude. However, the integration error of the numerical analysis is still approx. five times smaller than the half width of the confidence interval. 

\add{Results for the second property analyzed in this section are summarized in Table~\ref{tab:verify_simulation_prob}.} \remove{Table~3 summarizes t}\add{T}he probability that the standard demand is enabled, i.e., $\Phi_2:=m\Bigl({P_{d_1}^d}\Bigr) = 1$, at time $t' = \SI{8}{\hour}$ \add{is computed} for varying uniform and folded normal CDFs assigned to the competing general transitions $T^G_{d_{10}}$ and $T^G_{d_{12}}$\add{.} \remove{, where e}\add{E}ach column holds the results for one CDF. The firing time of $T^D_0$ is set to $\text{dat}_b=11$ h.
 Again, the results are computed for all three integration methods and compared to simulation. In this execution run, simulation was performed with the single confidence interval $\pm0.001$.
 Note that, the exact CDF used barely affects the runtime of the analysis.  The results are well supported by simulation and in most cases the computation time for all numerical approaches is considerably slower. The firing times for the deterministic transitions $T^D_{d_{10}}$ and $T^D_{d_{12}}$ are chosen such that there is no possibility to return to standard demand once it has changed. Hence, the probability of being in standard demand mode results in $T^D_{d_{10}}$ and $T^D_{d_{12}}$ not having fired in the interval $[\SI{0}{\hour},\SI{8}{\hour}]$. 

\begin{table}[tb]
    \centering
    \begin{tabularx}{\linewidth}{|l|p{1.5cm}|Y|Y|Y|Y|Y|Y|Y|Y|}
        \hline 
         $T^G_{d_{10}}, T^G_{d_{12}}$&&$\mathcal{U}\left( \SI{0}{\hour},\SI{10}{\hour}\right)$&$\mathcal{U}\left(\SI{6}{\hour},\SI{10}{\hour}\right)$&$\mathcal{N}\left(\mu\text{=}\SI{8}{\hour}, \sigma\text{=}\SI{1}{\hour}\right)$&$\mathcal{N}\left(\SI{7}{\hour}, \SI{1}{\hour}\right)$&$\mathcal{N}\left(\SI{7}{\hour}, \SI{2}{\hour}\right)$\\
         \hline
         \hline
         & dim & $4$ & $4$ & $4$ & $4$ & $4$ \\
         \hline
         \multirow{2}{*}{\shortstack[l]{\textbf{NUM} \\ \textit{intervals}}} 
         & $p\left(t',\Phi_2\right)$ & $0.040$ & $0.241$ & $0.249$ & $0.023$ & $0.070$ \\
         & $e\left(t',\Phi_2\right)$ & $1.3\cdot e^{-5}$ & $8.5\cdot e^{-5}$ & $5.5\cdot e^{-6}$ & $9.8\cdot e^{-6}$ & $2.4\cdot e^{-5}$  \\
         & comp. time & $\SI{4.16}{\second}$ & $\SI{4.04}{\second}$ & $\SI{8.17}{\second}$ & $\SI{8.21}{\second}$ & $\SI{7.62}{\second}$ \\
         \hline
         \multirow{2}{*}{\shortstack[l]{\textbf{NUM} \\ \textit{simplices}}} 
         & $p\left(t',\Phi_2\right)$ & $0.040$ & $0.242$ & $0.247$ & $0.024$ & $0.068$ \\
         & $e\left(t',\Phi_2\right)$ & $1.1\cdot e^{-6}$ & $6.4\cdot e^{-6}$ & $3.5\cdot e^{-6}$ & $1.5\cdot e^{-7}$ & $1.1\cdot e^{-6}$  \\
         & comp. time & $\SI{3.13}{\second}$ & $\SI{3.06}{\second}$ & $\SI{3.91}{\second}$ & $\SI{3.14}{\second}$ & $\SI{1.84}{\second}$ \\
         \hline \multirow{2}{*}{\shortstack[l]{\textbf{NUM} \\ \textit{polytopes}}} 
         & $p\left(t',\Phi_2\right)$ & $0.040$ & $0.241$ & $0.250$ & $0.023$ & $0.069$ \\
         & $e\left(t',\Phi_2\right)$ & $1.0\cdot e^{-6}$ & $6.5\cdot e^{-6}$ & $3.5\cdot e^{-6}$ & $1.5\cdot e^{-7}$ & $1.2\cdot e^{-6}$  \\
         & comp. time & $\SI{2.95}{\second}$ & $\SI{2.89}{\second}$ & $\SI{2.94}{\second}$ & $\SI{2.95}{\second}$ & $\SI{1.62}{\second}$ \\
         \hline
         \hline
         \multirow{2}{*}{\bf SIM} & $p\left(t',\Phi_2\right)$ & $0.041$ & $0.240$ & $0.250$ & $0.025$ & $0.072$ \\
         & comp. time\newline \tiny{(CI: $\pm 0.001$)} & $\SI{8.89}{\second}$ & $\SI{19.54}{\second}$ & $\SI{32.91}{\second}$ & $\SI{7.09}{\second}$ & $\SI{16.87}{\second}$ \\
         \hline
         
    \end{tabularx}
    
    \caption{Probability of a standard demand, i.e., $\Phi_2:=m\Bigl({P_{d_1}^d}\Bigr) = 1$, at time $t'=\SI{8}{\hour}$ for randomly distributed demand changing times with different distributions, computed numerically with different integration approaches and via simulation.  We state the computed probability $p(t',\Phi_2)$, the error from Monte Carlo integration $e(t', \Phi_2)$ and the computation time.}
    \label{tab:verify_simulation_prob}
\end{table}

\subsection{\add{Performance evaluation}}\label{ss:performance}

After focusing on the validation of the computed probabilities in the first part of the case study, the second part  further analyzes the performance of the different integration methods. We investigate the number of dimensions, the number of locations and the computation times, when checking whether $\Phi_3:= x\bigl({P_1^c}\bigr) > \SI{0}{\kWh}$ holds  for  increasing  time $t'$. 
%
In this setup the transition $T_0^G$ follows a folded normal distribution with mean $\mu=\SI{14}{\hour}$ and variance $\sigma=\SI{4}{\hour}$. Once the grid failed, it needs $22$ hours to be  repaired.  $T^G_{d_{10}}$ and $T^G_{d_{12}}$ follow a uniform distribution $\mathcal{U}(\SI{0}{\hour},\SI{2}{\hour})$ and the firing times of both $T^D_{d_{01}}$ and $T^D_{d_{21}}$ equals $\SI{5}{\hour}$.

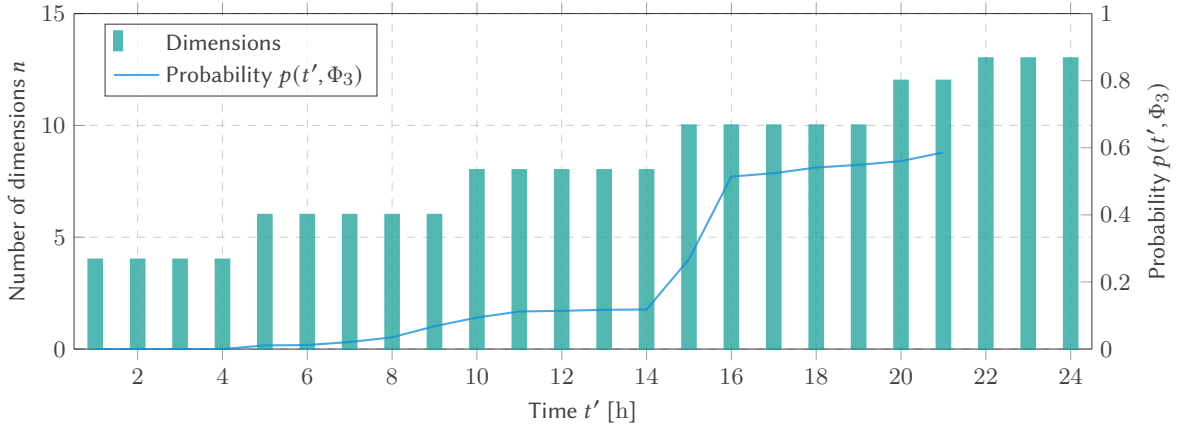
\begin{figure}
      \centering
      \input{images/dimensions.tex}
    \caption{ Number of dimensions of the state space for different times $t'$ (measured in hours), independent of the property and for $\tau_{\text{max}}=t'$ (left y-axis). Probability that $\Phi_3:= x\bigl({P_1^c}\bigr) > \SI{0}{\kWh}$ holds at time $t'$, computed with \textbf{NUM} \emph{intervals} (right y-axis). }
      \label{fig:dimensions}
\end{figure}

\begin{figure}
      \centering
      \input{images/locations.tex}
    \caption{Number of locations: (i) locations in the parametric location tree created up to $\tau_\text{max} = t'$ (black), (ii) the locations the system can be in at time $t'$  (measured in hours) (blue) and (iii)  the subset of locations satisfying $\Phi_3:= x\bigl({P_1^c}\bigr) > \SI{0}{\kWh}$ at time $t'$ (green).}
      \label{fig:locations}
\end{figure}
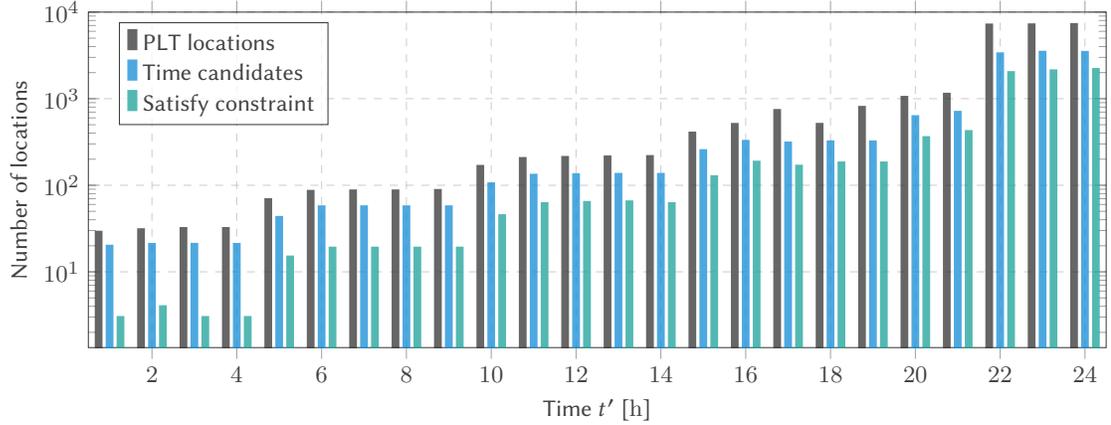

We evaluate the impact of a larger time on the size of the state space and on the computation times.  The state space is computed for $\tau_\text{max}=t'$, where a larger $t'$ results in more random variables  present in the system. 
Figure \ref{fig:dimensions}  indicates the number of dimensions for different values of $t'$. \add{Additionally, it shows the probability that extra cost have been accumulated, as the available battery charge was not sufficient.}
For $t' \leq \SI{5}{\hour}$ we obtain four dimensions, as all three general transitions can fire only once and the time also adds one dimension. In general, two dimensions are added every 5 hours, since the two general transitions that change the demand can both fire once more.
At $t' = \SI{22}{\hour}$ the number of dimensions increases to $13$, because  the grid can then fail twice. 

The \add{computed} probability $p(t',\Phi_3)$ is zero for $t'\leq \SI{4}{\hour}$ and then increases with the possibility of a grid failure. Note that, the steeper increase after $t'=\SI{14}{\hour}$ coincides with the mean of the grid failure distribution. The growing probability is however due to the battery level, which has been drained before and is no longer able to provide the facility with enough power in case of a grid failure.  For $t'=\SI{21}{\hour}$ the value $0.586$ is reached, which is the last probability that could be computed. For later time points neither of the integration methods is able to compute the corresponding probability in less than 2 hours.

The y-axis in Figure \ref{fig:locations} indicates the number of locations  (i) in the PLT created up to time $\tau_\text{max} = t'$, (ii) the system can be in at time $t'$, and  (iii) that satisfy $\Phi_3$ at $t'$. Similar to the number of dimensions, the number of locations also increases periodically. At $t' \geq \SI{22}{\hour}$ the number of locations rises, considerably.
Both, the total number of locations, i.e., the size of the state space, as well as the dimensionality of the model has a great impact on the computation time of the transient distribution, increase. While the number of dimensions directly depends on the number of random variables present in the system and is hard to reduce, the number of locations could be reduced by e.g., a property-guided exploration of the state space. 

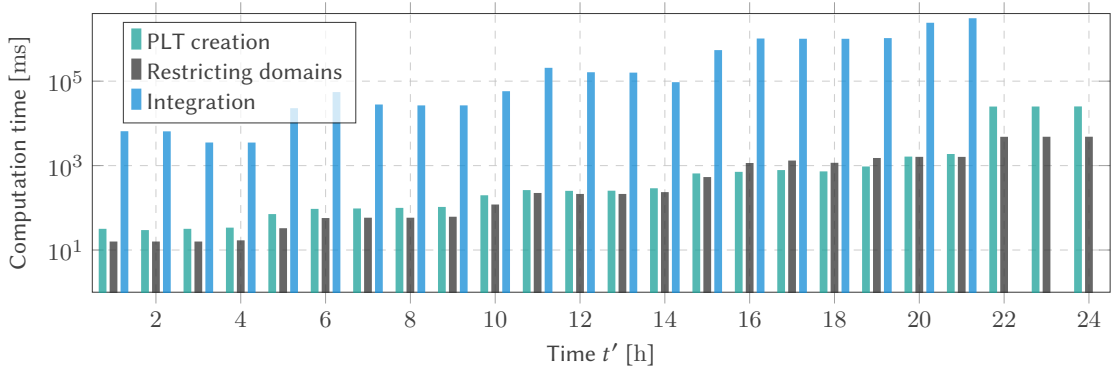
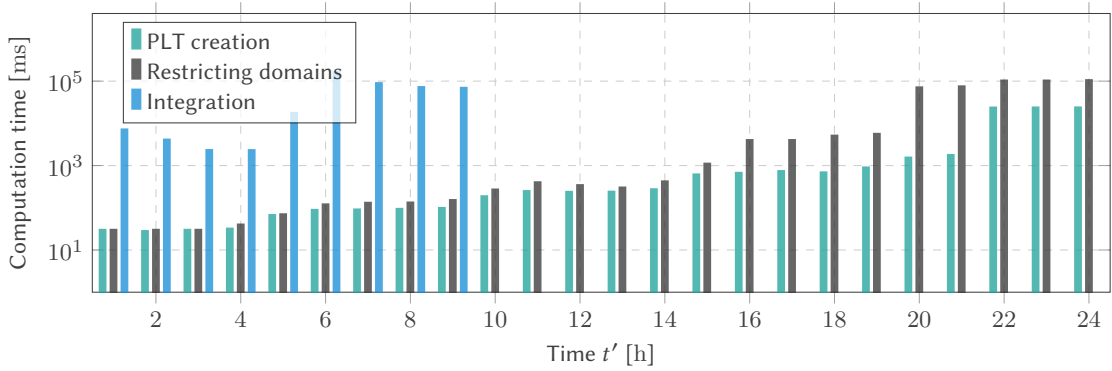
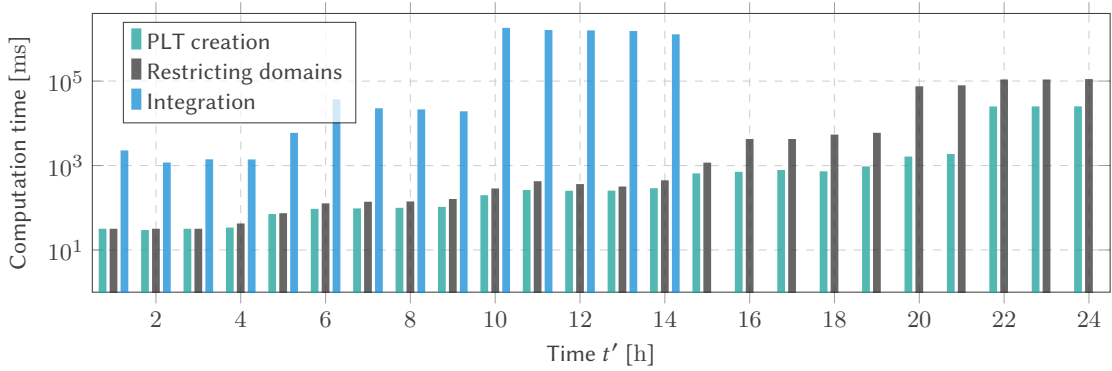
\begin{figure}[p]
    \centering
    \begin{subfigure}{\textwidth}
      \centering
      \input{images/computation_time.tex}
      \subcaption{Computation time of the \textbf{NUM} \textit{intervals} integration method for different time points.} \label{fig:time_interval}
    \end{subfigure}
    \vspace{1em}
    
    \begin{subfigure}{\textwidth}
      \centering
      \input{images/computation_time_simplices.tex}
      \subcaption{Computation time of the \textbf{NUM} \textit{simplices} integration method for different time points.} \label{fig:time_simplices}
    \end{subfigure}
    \vspace{1em}
    
    \begin{subfigure}{\textwidth}
      \centering
      \input{images/computation_time_polytopes.tex}
      \subcaption{Computation time of the \textbf{NUM} \textit{polytopes} integration method for different time points.} \label{fig:time_polytopes}
    \end{subfigure}
    \vspace{1em}
    
    \caption{Computation times for the probability that $\Phi_3:= x\bigl({P_1^c}\bigr) > \SI{0}{\kWh}$ holds at time $t'$ (measured in hours), with three different integration methods. In each plot, the overall computation time equals the sum of the three components: (i) PLT creation (green), (ii) Restricting domains (black) and (iii) Integration (blue).}
        \label{fig:time}
    \end{figure}




Figure~\ref{fig:time} separately indicates the time needed to (i) create the PLT up to time $t'$, (ii) compute the restricted domains and (iii) perform the integration for all presented integration methods w.r.t. constraint $\Phi_3$. For all integration methods, the time for integration takes an order of magnitude longer than creating the complete PLT and preparing the intervals for final integration. 
Note that, the time for creating the PLT is equal for all methods, since the creation is independent of the chosen method. Further, the time for restricting the domains is equal in  Figures~\ref{fig:time_simplices} and \ref{fig:time_polytopes}, as both integration approaches rely on the same geometric operations to prepare the intervals for integration.  Transformation to standard simplices, as well as all operations performed for direct integration are included into the total integration time. 
Comparing the computation times indicated in Figure~\ref{fig:time_interval} with Figures~\ref{fig:time_simplices} and \ref{fig:time_polytopes}, shows that geometrically restricting the domains takes longer than restricting the intervals arithmetically.  However, the geometric approach is more general and enables model checking more complex properties. 
 
Similar to the results shown in Table~\ref{tab:verify_simulation_time}, the \textbf{NUM} \textit{simplices} method was unable to terminate for  time $t'$ larger than $\SI{9}{\hour}$ and the \textbf{NUM} \textit{polytopes} method for $t'$ larger than $\SI{14}{\hour}$. Only for a time $t' \geq \SI{22}{\hour}$, the \textbf{NUM} \textit{intervals} method was unable to complete within 2 hours. Hence, the limiting factor of the presented approach clearly are the number of dimensions and the number of required triangulations and integrations.  

Summarizing, we can conclude for lower dimensions that \textbf{NUM} \textit{simplices} and \textbf{NUM} \textit{polytopes} perform better than \textbf{NUM} \textit{intervals}, which becomes more efficient for larger dimensions. This can be explained by the computational and storage complexity of the geometrical operations. Both, computing the Delauney triangulation  for \textbf{NUM} \textit{simplices} and the bounding box for \textbf{NUM} \textit{polytopes} require the transformation of the hyperplane-representation into the vertex representation for polytopes\add{.} \remove{, which can be exponential in the state-space dimension.} 

\paragraph{\add{Reproducability.}}
All computations have been performed on a MacBook Pro with $\SI{2.5}{\giga\Hz}$ Inter Core i7 and $\SI{16}{\giga\byte}$ of memory. 
\add{The source code for the tool \hpnmg can be accessed at \texttt{\url{https://zivgitlab.uni-muenster.de/ag-sks/tools/hpnmg}}.
The tags \texttt{num-simplices} and \texttt{num-polytopes} lead to the different versions using the corresponding integration method. Furthermore, scripts for reproducing the above results  can be obtained from  \texttt{\url{https://zivgitlab.uni-muenster.de/ag-sks/tools/misc/execution-scripts/2020-09-qest-special-issue}}.}

\subsection{\add{CPU and memory usage}} \label{sec:CPU-memory}


\add{As shown in the previous section, both versions of the geometric computation are more restricted w.r.t the number of dimensions than the interval integration method. This  is expected to be a consequence of the geometric operations necessary for computing the restricted domains and the inherent transformations from the $\mathcal{H}$-representation to $\mathcal{V}$-representation of polytopes and vice-versa, which is known to be a memory-intensive task. Note that, these changes of representation are not required for the interval method. However, the interval method can only be used to compute transient probabilities and is not suited for model checking more complex properties.} 

\add{To further analyze memory and CPU usage for the geometric methods, we use the tool \psrecord}\footnote{\add{\url{https://github.com/astrofrog/psrecord}}}. \add{We record the CPU and memory activity for a set of experiments with both geometric integration approaches. On every experiment, \psrecord is directly attached to the specific process and only measures the accumulated process-specific memory and CPU activity. }

\begin{figure}[p]
    \centering
    %
    \begin{subfigure}{0.48\textwidth}
        \centering
        \includegraphics[width=\textwidth]{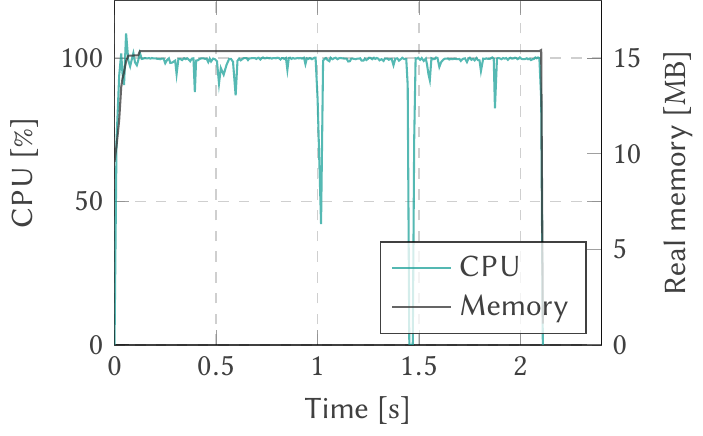}
        \subcaption{\add{\textbf{NUM} \textit{polytopes} integration method for $t'=\SI{3}{\hour}$.}} 
        \label{fig:num-poly-3}
    \end{subfigure}
    %
    \begin{subfigure}{0.48\textwidth}
        \centering
        \includegraphics[width=\textwidth]{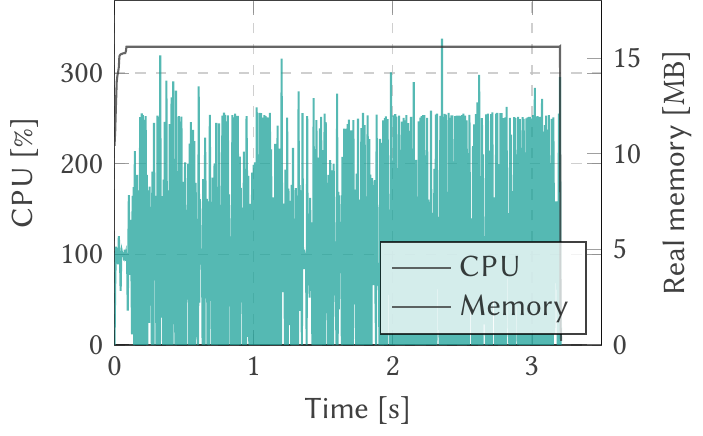}
        \subcaption{\add{\textbf{NUM} \textit{simplices} integration method for $t'=\SI{3}{\hour}$.}} 
        \label{fig:num-sim-3}
    \end{subfigure}
    \vspace{1em}
    
    %
    \begin{subfigure}{0.48\textwidth}
        \centering
        \includegraphics[width=\textwidth]{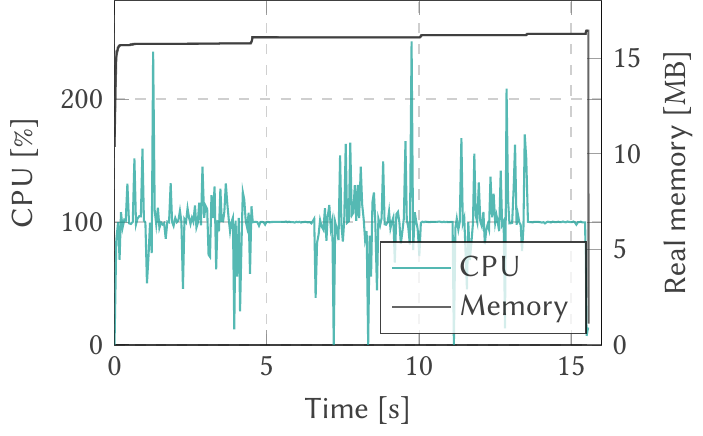}
        \subcaption{\add{\textbf{NUM} \textit{polytopes} integration method for $t'=\SI{9}{\hour}$.}} 
        \label{fig:num-poly-9}
    \end{subfigure}
    %
    \begin{subfigure}{0.48\textwidth}
        \centering
        \includegraphics[width=\textwidth]{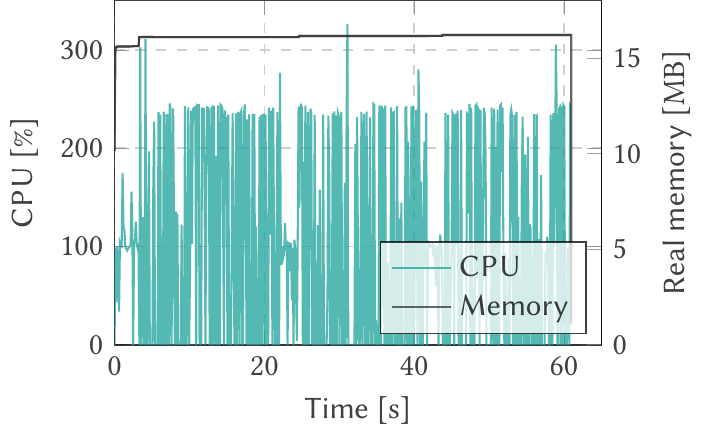}
        \subcaption{\add{\textbf{NUM} \textit{simplices} integration method for $t'=\SI{9}{\hour}$.}} 
        \label{fig:num-sim-9}
    \end{subfigure}
    \vspace{1em}
    
    %
    \begin{subfigure}{0.48\textwidth}
        \centering
        \includegraphics[width=\textwidth]{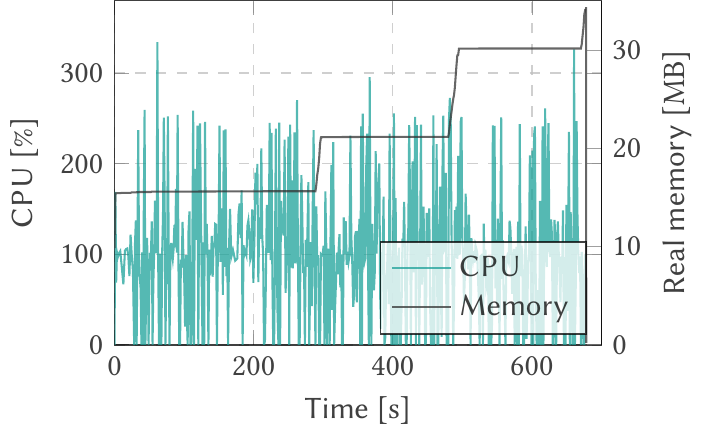}
        \subcaption{\add{\textbf{NUM} \textit{polytopes} integration method for $t'=\SI{10}{\hour}$.}} 
        \label{fig:num-poly-10}    
    \end{subfigure}    
    %
    \begin{subfigure}{0.48\textwidth}
        \centering
        \input{images/memory/constraints}
        \subcaption{\add{Constraints required for the computation of restricted domains and number of dimensions for $t'=\SI{1}{\hour},\dots,\SI{14}{\hour}$.}} 
        \label{fig:constraints}
    \end{subfigure}
    
    \caption{
    \add{CPU and memory usage for the integration methods \textbf{NUM} \textit{polytopes} and \textbf{NUM} \textit{simplices} for three different values for $t'$ (\subref{fig:num-poly-3}-\subref{fig:num-poly-10}) and the average number of constraints required for computing the transient probability at time $t'$ and the number of dimensions  (\subref{fig:constraints}).
    Note that, for each of the plots the same amount of points was used to plot the CPU and memory usage, even though the $x$-axis have different maximum values for the time (depending on the total computation time).
    }}
    \label{fig:cpu}
\end{figure}

\add{The measurements shown in Figure~\ref{fig:cpu} correspond to the case study setting also discussed in Figure~\ref{fig:time}. The plots from  Figures \ref{fig:num-poly-3}-\ref{fig:num-poly-10} show CPU and memory usage for the computations required for $t'=\SI{3}{\hour}, \SI{9}{\hour}$ and $\SI{10}{\hour}$, where the computation time (in seconds) is provided on the $x$-axis and CPU and memory usage are depicted on the respective $y$-axes. In contrast, Figure~\ref{fig:constraints} shows the number of constraints resulting from the computation of the restricted domains, as required for the computation of the transient probability at different times $t'$, as indicated on the $x$-axis, as well as the resulting number of dimensions of the state space. }

\add{ The plots in Figures~\ref{fig:num-poly-3} and \ref{fig:num-sim-3} show that for $t'=\SI{3}{\hour}$ memory usage is constant around $\SI{15.4}{\mega\byte}$ for both integration methods, while CPU usage varies considerably for \textbf{NUM} \textit{simplices} between $\SI{0}{\percent}$ and $\SI{300}{\percent}$ and only rarely drops from the usual $\SI{100}{\percent}$ for the \textbf{NUM} \textit{polytopes} method. 
Observing the plots for $t'=\SI{9}{\hour}$ in Figures~\ref{fig:num-poly-9} and \ref{fig:num-sim-9}, one can see that memory usage slightly increases over the runtime of the computation for both methods. Also CPU usage starts to vary for \textbf{NUM} \textit{polytopes} and even more so for $t'=\SI{10}{\hour}$ (Figure~\ref{fig:num-poly-10}). The transient probability for $t'=\SI{10}{\hour}$ could only be obtained from the \textbf{NUM} \textit{polytopes} method. The corresponding memory usage for that computation clearly shows an increased memory usage over the duration of the computation from $\SI{15 }{\mega\byte}$ to $\SI{30.1}{\mega\byte}$ in three steps with an additional peak at the end of the computation around $\SI{34}{\mega\byte}$. More detailed measurements have shown that a new step is introduced when integration is called for an additional candidate location. The  geometric computations required to remove the time dimension from the representation of the convex polytopes accounts for the additional memory usage.}

\add{
Overall, the CPU usage for \textbf{NUM} \textit{simplices} is considerably higher than for \textbf{NUM} \textit{polytopes}, while the memory usage is almost similar for both methods up to $t'=\SI{9}{\hour}$. Hence, we expect that the combination of high CPU usage with the highly increased memory usage, as observed in Figure~\ref{fig:num-poly-10}, is the reason why \textbf{NUM} \textit{simplices} is unable to compute  results for $t'\geq \SI{10}{\hour}$.}

\add{A direct dependency can be observed between the number of dimensions and constraints, as illustrated in Figure~\ref{fig:constraints}, and the corresponding memory and CPU usage. We have illustrated CPU and memory usage over the full computation time for different times $t'$ resulting in $8$, $14$ and $20$ constraints. }

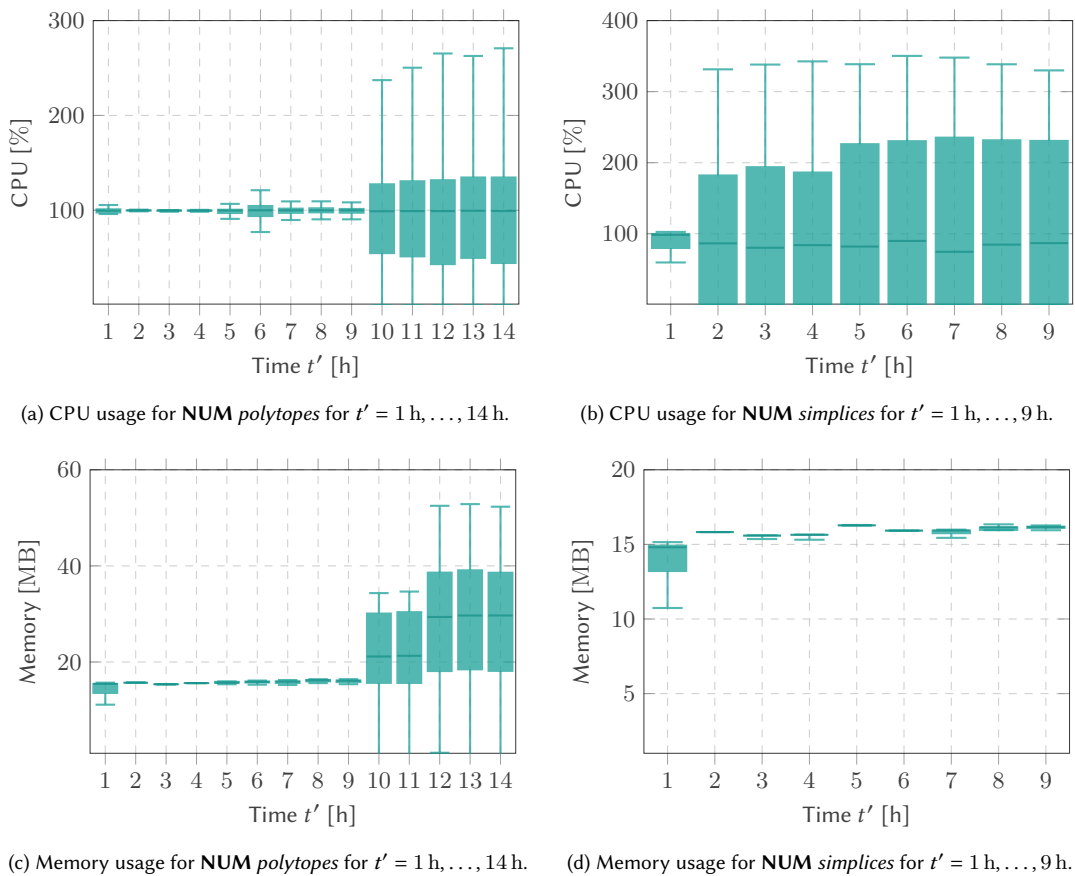
\begin{figure}
    \centering
    %
    \begin{subfigure}{0.48\textwidth}
        \centering
        \input{images/memory/cpu_quantile_poly}
        \subcaption{\add{CPU usage for \textbf{NUM} \textit{polytopes} for $t'=\SI{1}{\hour},\dots,\SI{14}{\hour}$.}} 
        \label{fig:cpu_quan_poly}
    \end{subfigure}
    %
    \begin{subfigure}{0.48\textwidth}
        \centering
        \input{images/memory/cpu_quantile_sim}
        \subcaption{\add{CPU usage for \textbf{NUM} \textit{simplices} for $t'=\SI{1}{\hour},\dots,\SI{9}{\hour}$.}}
        \label{fig:cpu_quan_sim}
    \end{subfigure}
    \vspace{1em}
     
    %
    \begin{subfigure}{0.48\textwidth}
        \centering
        \input{images/memory/mem_quantile_poly}
        \subcaption{\add{Memory usage for \textbf{NUM} \textit{polytopes} for $t'=\SI{1}{\hour},\dots,\SI{14}{\hour}$.}}
        \label{fig:mem_quan_poly}
    \end{subfigure}
    %
    \begin{subfigure}{0.48\textwidth}
        \centering
        \input{images/memory/mem_quantile_sim}
        \subcaption{\add{Memory usage for \textbf{NUM} \textit{simplices} for $t'=\SI{1}{\hour},\dots,\SI{9}{\hour}$.}}
        \label{fig:mem_quan_sim}
    \end{subfigure}
    %
    \caption{\add{Boxplots illustrating the CPU (\subref{fig:cpu_quan_poly} and \subref{fig:cpu_quan_sim}) and memory usage (\subref{fig:mem_quan_poly} and \subref{fig:mem_quan_sim}) for the two integration methods \textbf{NUM} \textit{polytopes} and \textbf{NUM} \textit{simplices} for increasing values of $t'$.
    The boxplots show the median, the lower and upper quartile of the values as well as the lower and upper whiskers, indicating the distribution of the measured values.
    }}
    \label{fig:quantiles}
\end{figure}

 \add{For a more complete comparison of memory and CPU usage, Figure~\ref{fig:quantiles} shows boxplots for memory and CPU usage for both geometric integration methods for all times $t'$, which terminate within $\SI{2}{\hour}$. This results in different scales on the $x$-axes of the plots. Note that, due to different measurements also the scales on $y$-axes differ. }

\add{ Figures~\ref{fig:cpu_quan_poly} and \ref{fig:cpu_quan_sim} show the distribution of CPU usage for varying values of $t'$.  
 While the median remains stable around $\SI{100}{\percent}$ for  \textbf{NUM} \textit{polytope}, it varies slightly for  \textbf{NUM} \textit{simplices}. It can be seen that in Figure~\ref{fig:cpu_quan_poly} CPU usage varies with an  increasing number of constraints. Especially when reaching $20$ constraints (i.e., for $t'\geq \SI{10}{\hour}$), we see that half of the CPU measurements now lie approximately between $\SI{50}{\percent}$ and $\SI{135}{\percent}$, resulting in an interquartile range of $\SI{85}{\percent}$. The whiskers indicate those measurements which are just included in  the quartiles extended by  $1.5$ times the interquartile range.
For $20$ constraints the lower whiskers are at $\SI{0}{\percent}$ CPU usage and the upper whiskers vary between $\SI{237}{\percent}$ and  $\SI{270}{\percent}$. 
In contrast, Figure~\ref{fig:cpu_quan_sim} shows that the interquartile ranges for the CPU measurements with \textbf{NUM} \textit{simplices} are considerably larger, starting at time $t'=\SI{1}{\hour}$. The effect of the constraints is less visible in this boxplot, than for \textbf{NUM} \textit{polytopes}.}
 
\add{ Considering the boxplots for memory usage for both integration methods, as illustrated in Figures~\ref{fig:mem_quan_poly} and \ref{fig:mem_quan_sim}, one can again observe a direct relationship between the number of constraints and the variance in the measurements. For less than $20$ constraints (i.e., $t'\leq \SI{9}{\hour}$) the interquartile range is around $\SI{0.5}{\mega\byte}$ with an outlier of $\SI{2}{\mega\byte}$ for time $t'=\SI{1}{\hour}$. For more constraints, the interquartile range increases to $\SI{15}{\mega\byte}$ ($t'=\SI{10}{\hour}$ and $t'=\SI{11}{\hour}$) and $\SI{20}{\mega\byte}$ ($t'=\SI{12}{\hour}, \SI{13}{\hour}, \SI{14}{\hour}$), while the median also increases to approximately $\SI{21}{\mega\byte}$ and $\SI{29.5}{\mega\byte}$, respectively. 
 Taking into account the different axes, Figure~\ref{fig:mem_quan_sim} illustrates similar quartiles for  \textbf{NUM} \textit{simplices}.}

 \add{These results can be explained as follows: 
 When applying  \textbf{NUM} \textit{simplices} the integration is carried out by GSL, which samples from the unit cube and internally compares the sample with the bounds obtained from triangulation. 
 Note that, integration is called once for every simplex, which each results in a peak in CPU usage. Since the integration is solely carried out by GSL, we did not further look into the performance of these calls. }
 
 \add{In contrast, \textbf{NUM} \textit{polytopes} calls the integration via GSL once for each pre-computed bounding box and then relies on  \hypro to geometrically test containment for each sample. Further detailed measurements have shown, that for a parameter setting resulting in $8$ constraints, the CPU is mainly utilized for containment testing. 
 With an increasing number of constraints the CPU and memory usage required for geometric operations and the corresponding change of representation becomes inhibitive. It is possible that the combination of third-party libraries results in additional overhead, which could be reduced in the future.}

    \FloatBarrier

%% file: images/batterymodel.tex
%
%
%
%
%
%

	\begin{tikzpicture}[scale=0.85,
	font=\sffamily,	
	every text node part/.style={align=center},
	cPlace/.style={thick,circle, draw, double distance=0.5mm, minimum size = 1cm},
	dPlace/.style={thick,circle, draw, minimum size = 1.1cm},
	dTrans/.style={thick,rectangle, draw, minimum size = 0.5cm, fill = gray, minimum height = 1.1cm},
	iTrans/.style={thick,rectangle, draw, minimum size = 0.15cm, fill = black, minimum height = 1.1cm},
	gTrans/.style={thick,rectangle, draw, minimum size = 0.5cm, minimum height = 1.1cm},
	cTrans/.style={thick,rectangle, draw, double distance=0.5mm, minimum size = 0.4cm, minimum height = 1cm},
	dyTrans/.style={thick,rectangle, draw, double distance=0.5mm, fill=black, minimum size = 0.4cm, minimum height = 1cm},
	cArc/.style={decoration={markings, mark=at position 1 with {\arrow[scale=1.5]{open triangle 60}}},
		double distance=0.5mm, 
		shorten >= 3.199mm, 
		shorten <= 0.39mm, 
		preaction = {decorate},
		thick},
	dArc/.style={thick,> = stealth', ->, thick},
	gArc/.style={thick,> = triangle 60, <->, thick}, 
	gcArc/.style={thick,> = triangle 60, <->, thick, shorten <= 0.25mm}, 
	gccArc/.style={thick,> = triangle 60, <->, thick, shorten <= 0.25mm, shorten >= 0.25mm}, 
    iArc/.style={thick,-o, thick, shorten <= 0.25mm} 
	]

	\draw[draw=none] (0.5,2.5) rectangle (16.53,-7.75);
	
	\node (pd1) [dPlace, label=above: $P^d_1$] at (1.25,-1) {};
	\node (pd0) [dPlace, label=above: $P^d_0$] at (6.25,-1) {\CIRCLE};
	\node (tg0) [gTrans, label={above: $T_0^G$}, label={}] at (3.75,-2.25)	{};
	\node (td0) [dTrans, label=above: $T_0^D$, label={below:\footnotesize $\mathit{dat}_b$}] at (3.75,0.25) {};
	\draw[dashed, thick, rounded corners] ($(pd1)+(-1,+3)$) rectangle ($(pd0)+(+2,-3)$);

	\draw [dArc, bend left=20] (pd0) to (tg0);
	\draw [dArc, bend left=20] (tg0) to (pd1);
	\draw [dArc, bend left=20] (pd1) to (td0);
	\draw [dArc, bend left=20] (td0) to (pd0);
	
	\node (td0) [dyTrans, label={above: $T^\mathit{Dy}_0$}, label={below:\footnotesize $\max\left(\left(P_A - \sum d_i\right), 0\right) $}] at (10,-1) {};
	\node (pc0) [cPlace, label=above: $P^c_0$, label={below:\footnotesize $B$}] at (12.5,-1) {};
	\node (td1) [dyTrans, label=above: $T^\mathit{Dy}_1$, label={below:\footnotesize $\max\left(\left(\sum d_i - P_A\right), 0\right) $}] at (15,-1) {};
	
	\draw [cArc] (td0) to (pc0);
	\draw [cArc] (pc0) to (td1);
	
	\node (ti0) [iTrans, label=above:$T^I_0$\hspace{1em}] at (10,-4.75) {};
	\node (ti1) [iTrans, label=above:$T^I_1$] at (15,-4.75) {};
	\node (pd2) [dPlace, label=above:$P^d_2$] at (12.5,-3.5) {\CIRCLE};
	\node (pd3) [dPlace, label=above:$P^d_3$] at (12.5,-6) {};
	
	\draw [dArc, bend right=20] (ti0) to (pd3);
	\draw [dArc, bend right=20] (pd3) to (ti1);
	\draw [dArc, bend right=20] (ti1) to (pd2);
	\draw [dArc, bend right=20] (pd2) to (ti0);	
	
	\node (td2) [dyTrans, label=above:$T^\mathit{Dy}_2$, label={below:\footnotesize $\max\left(\left(\sum d_i - P_A\right), 0\right) $}] at (6.25,-6) {};
	\node (pc1) [cPlace, label=above:$P^c_1$, label={below:\footnotesize $\infty$}] at (3.75,-6){};
	
	\draw [cArc] (td2) to (pc1);
	
	\draw [gcArc] (td0) to (pd0); 
	\draw [gArc] (pc0) to node [left, draw=none] {\footnotesize $>0$}(ti1.north west); 
	\draw [iArc] (pc0) to node [right, draw=none] {\footnotesize $>0$
    } (ti0.north east); 
	\draw [gcArc] (td2) to (pd3); 
	\end{tikzpicture}	

%% file: images/demands.tex
	
	\begin{tikzpicture}[scale=0.85,
	font=\sffamily,	
	every text node part/.style={align=center},
	cPlace/.style={thick,circle, draw, double distance=0.5mm, minimum size = 1cm},
	dPlace/.style={thick,circle, draw, minimum size = 1.1cm},
	dTrans/.style={thick,rectangle, draw, minimum size = 0.5cm, fill = gray, minimum height = 1.1cm},
	iTrans/.style={thick,rectangle, draw, minimum size = 0.15cm, fill = black, minimum height = 1.1cm},
	gTrans/.style={thick,rectangle, draw, minimum size = 0.5cm, minimum height = 1.1cm},
	cTrans/.style={thick,rectangle, draw, double distance=0.5mm, minimum size = 0.4cm, minimum height = 1cm},
	dyTrans/.style={thick,rectangle, draw, double distance=0.5mm, fill=black, minimum size = 0.4cm, minimum height = 1cm},
	cArc/.style={decoration={markings, mark=at position 1 with {\arrow[scale=1.5]{open triangle 60}}},
		double distance=0.5mm, 
		shorten >= 3.199mm, 
		shorten <= 0.39mm, 
		preaction = {decorate},
		thick},
	dArc/.style={thick,> = stealth', ->, thick},
	gArc/.style={thick,> = triangle 60, <->, thick}, 
	gcArc/.style={thick,> = triangle 60, <->, thick, shorten <= 0.25mm}, 
	gccArc/.style={thick,> = triangle 60, <->, thick, shorten <= 0.25mm, shorten >= 0.25mm} 
	]
	
	\draw[draw=none] (0.7,1.5) rectangle (8.75,-8.75);
	
		\node (tf_d0) [cTrans, label=above:$T^F_{d_0}$, label={below: \footnotesize $d_0$}] at (1.25,0) {};
	\node (tf_d1) [cTrans, label=above:$T^F_{d_1}$, label={below: \footnotesize $d_1$}] at (1.25,-3.75) {};
	\node (tf_d2) [cTrans, label=above:$T^F_{d_2}$, label={below: \footnotesize $d_2$}] at (1.25,-7.5) {};
	
	\node (pd_d0) [dPlace, label=above:$P^d_{d_0}$] at (5.75,0) {};
	\node (pd_d1) [dPlace, label=above:$P^d_{d_1}$] at (5.75,-3.75) {\CIRCLE};
	\node (pd_d2) [dPlace, label=above:$P^d_{d_2}$] at (5.75,-7.5) {};
	
	\node (td_d0_1) [dTrans, label=above:$T^D_{d_{01}}$, label={below: \footnotesize $\mathit{at}_1 $}] at (7.5,-1.875) {};
	\node (td_d1_2) [gTrans, label=above:$T^G_{d_{12}}$, label={below: \footnotesize $\mathit{at}_2 $}] at (7.5,-5.625) {};
	
	\node (td_d1_0) [gTrans, label=above:$T^G_{d_{10}}$, label={below: \footnotesize $\mathit{dat}_1 $}] at (4,-1.875) {};
	\node (td_d2_1) [dTrans, label=above:$T^D_{d_{21}}$, label={below: \footnotesize $\mathit{dat}_2 $}] at (4,-5.625) {};

	\draw [dArc] (pd_d0) to (td_d0_1);
	\draw [dArc] (td_d0_1) to (pd_d1);
	\draw [dArc] (pd_d1) to (td_d1_2);
	\draw [dArc] (td_d1_2) to (pd_d2);
	\draw [dArc] (pd_d2) to (td_d2_1);
	\draw [dArc] (td_d2_1) to (pd_d1);
	\draw [dArc] (pd_d1) to (td_d1_0);
	\draw [dArc] (td_d1_0) to (pd_d0);
	
	\draw [gArc] (tf_d0) to (pd_d0);
	\draw [gArc] (tf_d1) to (pd_d1);
	\draw [gArc] (tf_d2) to (pd_d2);
	
	\end{tikzpicture}	
	

%% file: images/dimensions.tex
\begin{tikzpicture}[font=\sffamily, fill opacity = 0.75, opacity = 0.75]

\pgfplotsset{ width=\textwidth,
height=0.4\textwidth,
xlabel=Time $t'$ [\si{\hour}], 
xmajorgrids=true,
        xmin=0.5,
        xmax=24.5,  
        legend pos=north west
}
    \begin{axis}[colormap/viridis,
        ybar,        
        ylabel=Number of dimensions $n$, 
        ymajorgrids=true,         
        grid style=dashed,
        bar width=0.18cm,
        ymin=0,
        ymax=15,
        axis y line*=left,
        ,]
      \addplot+[ybar,ybar legend] [color=valuecolour, fill, thick, name path=A] table[x=maxtime, y=dimension] {images/locations.dat};\label{plot_one}

    \end{axis}
    
     \begin{axis}[colormap/viridis,
         ylabel= {}{Probability $p(t', \Phi_3)$},      
         ymajorgrids=false, 
     bar width=0.18cm,
         ymin=0,
         ymax=1,
         axis y line*=right,
         axis x line=none,
         every axis legend/.append style={nodes={right}},
         ,]
         \addlegendimage{/pgfplots/refstyle=plot_one, ybar, fill, thick}\addlegendentry{Dimensions}
       \addplot[color=keycolour, thick, name path=A] coordinates{
    (1,0)
    (2,0)
    (3,0)
    (4,0)
    (5,0.011)
    (6,0.012)
    (7,0.021)
    (8,0.035)
    (9, 0.068)
    (10,0.094)
    (11,0.112)
    (12,0.114)
    (13,0.117)
    (14,0.118)
    (15,0.267823)
    (16,0.51433)
    (17,0.52411)
    (18, 0.5411)
    (19,0.5492)
    (20,0.560172)
    (21,0.58632)
}; \addlegendentry{{}{Probability $p(t', \Phi_3)$}};

     \end{axis}
\end{tikzpicture}

%% file: images/locations.tex
\begin{tikzpicture}[font=\sffamily, fill opacity = 0.75, opacity = 0.75]
    \begin{axis}[colormap/viridis,
        ybar,
        xlabel={Time $t'$ [\si{\hour}]}, 
        ylabel=Number of locations, 
        legend pos=north west,
        ymajorgrids=true, 
        xmajorgrids=true,
        grid style=dashed,
        bar width=0.075cm,
        ymax=10000,
        ymode=log,
        xmin=0.5,
        xmax=24.5,
        width=\textwidth,
        height=0.4\textwidth,
        ,legend cell align={left}]
      \addplot[color=darkcolour, fill,  thick, name path=A] table[x=maxtime, y=total] {images/locations.dat};
      \addplot[color=keycolour,fill,   thick, name path=B] table[x=maxtime, y=candidates] {images/locations.dat};
      \addplot[color=valuecolour,fill, thick, name path=C] table[x=maxtime, y=satisfy] {images/locations.dat};
      \legend{PLT locations, Time candidates, Satisfy constraint};


    \end{axis}
\end{tikzpicture}

%% file: images/computation_time.tex
\pgfplotsset{compat=1.14}

\begin{tikzpicture}[font=\sffamily, fill opacity = 0.75, opacity = 0.75]

    \begin{axis}[colormap/viridis,
        ybar,
        xlabel={Time $t'$ [\si{\hour}]},
        ylabel={Computation time [\si{\milli\second}]}, 
        legend pos=north west,
        ymajorgrids=true, 
        xmajorgrids=true,
        grid style=dashed,
        bar width=0.075cm,
        ymin=1,
        ymax=4000000,
        ymode=log,
        xmin=0.5,
        xmax=24.5,
        width=\textwidth,
        height=0.35\textwidth,
        ,legend cell align={left}]
        \addplot[color=valuecolour,fill=valuecolour, thick, name path=C] table[x=maxtime, y=t_plt] {images/computation_time.dat};
        \addplot[color=darkcolour, fill=darkcolour, thick, name path=B] table[x=maxtime, y=t_intervals] {images/computation_time.dat};
      \addplot[ color=keycolour, fill=keycolour, thick, name path=A] table[x=maxtime, y=t_integration] {images/computation_time.dat};    
      
      \legend{PLT creation, Restricting domains, Integration};

    \end{axis}
\end{tikzpicture}

%% file: images/computation_time_simplices.tex
\pgfplotsset{compat=1.14}

\begin{tikzpicture}[font=\sffamily, fill opacity = 0.75, opacity = 0.75]

    \begin{axis}[colormap/viridis,
        ybar,
        xlabel={Time $t'$ [\si{\hour}]},
        ylabel={Computation time [\si{\milli\second}]}, 
        legend pos=north west,
        ymajorgrids=true, 
        xmajorgrids=true,
        grid style=dashed,
        bar width=0.075cm,
        ymin=1,
        ymax=4000000,
        ymode=log,
        xmin=0.5,
        xmax=24.5,
        width=\textwidth,
        height=0.35\textwidth,
        ,legend cell align={left}]
        \addplot[color=valuecolour,fill=valuecolour, thick, name path=C] table[x=maxtime, y=t_plt] {images/computation_time_simplices.dat};
        \addplot[color=darkcolour, fill=darkcolour, thick, name path=B] table[x=maxtime, y=t_intervals] {images/computation_time_simplices.dat};
      \addplot[ color=keycolour, fill=keycolour, thick, name path=A] table[x=maxtime, y=t_integration] {images/computation_time_simplices.dat};

      \legend{PLT creation, Restricting domains, Integration};


    \end{axis}
\end{tikzpicture}

%% file: images/computation_time_polytopes.tex
\pgfplotsset{compat=1.14}

\begin{tikzpicture}[font=\sffamily, fill opacity = 0.75, opacity = 0.75]

    \begin{axis}[colormap/viridis,
        ybar,
        xlabel={Time $t'$ [\si{\hour}]},
        ylabel={Computation time [\si{\milli\second}]}, 
        legend pos=north west,
        ymajorgrids=true, 
        xmajorgrids=true,
        grid style=dashed,
        bar width=0.075cm,
        ymin=1,
        ymax=4000000,
        ymode=log,
        xmin=0.5,
        xmax=24.5,
        width=\textwidth,
        height=0.35\textwidth,
        ,legend cell align={left}]
        \addplot[color=valuecolour,fill=valuecolour, thick, name path=C] table[x=maxtime, y=t_plt] {images/computation_time_polytopes.dat};
        \addplot[color=darkcolour, fill=darkcolour, thick, name path=B] table[x=maxtime, y=t_intervals] {images/computation_time_polytopes.dat};
      \addplot[ color=keycolour, fill=keycolour, thick, name path=A] table[x=maxtime, y=t_integration] {images/computation_time_polytopes.dat};

      \legend{PLT creation, Restricting domains, Integration};


    \end{axis}
\end{tikzpicture}

%% file: images/memory/constraints.tex
\pgfplotsset{compat=1.12}
\begin{tikzpicture}[font=\sffamily, fill opacity = 0.75, opacity = 0.75]

\pgfplotsset{ 
legend pos=south east,
legend cell align={left},
}

    \begin{axis}[colormap/viridis,
        ybar,
bar width=0.075cm,
bar shift = -0.065cm,
xlabel={Time $t'$ [\si{\hour}]},
        ylabel={Avg. number of constraints}, 
        ymajorgrids=true, 
        xmajorgrids=true,
        grid style=dashed,
        ymin=0,
        ymax=22,
        xmin=0.6,
        xmax=14.4,
        width=0.9\textwidth,
        height=0.7\textwidth,
        legend cell align={left},
        ]
        
        \addplot [color=keycolour, fill=keycolour, thick, ybar, ybar legend] table {images/memory/data/avg_constraints.dat};\label{constraints}

    \end{axis}
    
\begin{axis}[colormap/viridis,
axis y line*=right,axis x line=none,
        ylabel=Number of dimensions $n$, 
        ymin=0,
        ymax=11,
        xmin=0.6,
        xmax=14.4,
bar width=0.075cm,
bar shift = 0.065cm,
width=0.9\textwidth,
        height=0.7\textwidth,
]
    
\addlegendimage{/pgfplots/refstyle=constraints}\addlegendentry{Constraints}
\addplot[ybar,ybar legend] [color=darkcolour, fill, thick, name path=A] table[x=maxtime, y=dimension] {images/memory/data/dimensions.dat};
\label{dimensions}
\addlegendentry{Dimensions}
    
    \end{axis}
    
\end{tikzpicture}

%% file: images/memory/cpu_quantile_poly.tex
\pgfplotsset{compat=1.12}

\begin{tikzpicture}[font=\sffamily, fill opacity = 0.75, opacity=0.75]

\pgfplotsset{
boxplot/every box/.style={valuecolour, fill=valuecolour, semithick},
boxplot/every median/.style={valuecolour!90!black,  thick},
boxplot/every whisker/.style={valuecolour, thick},
scale only axis=true,
width=0.78\textwidth,
height=0.52\textwidth,
xtick={1,2,3,4,5,6,7,8,9,10,11,12,13,14}
}

    \begin{axis}[colormap/viridis,
        ybar,
        xlabel={Time $t'$ [h]},
        ylabel={CPU [\si{\percent}]}, 
        legend pos=north west,
        ymajorgrids=true, 
        xmajorgrids=true,
        grid style=dashed,
        bar width=0.075cm,
        ymin=1,
        ymax=300,
        xmin=0.5,
        xmax=14.5,
        legend cell align={left},
        boxplot,
        boxplot/draw direction = y,
        table/y=m,
        ]
        
        \addplot+[
boxplot prepared={lower whisker=96, lower quartile=97.35, median=99.5, upper quartile=101.225, upper whisker=105.5}
        ] coordinates {}; 
        \addplot+[
boxplot prepared={lower whisker=98.8, lower quartile=99.5, median=99.8, upper quartile=100, upper whisker=100.6}
        ] coordinates {}; 
        \addplot+[
boxplot prepared={lower whisker=98.4, lower quartile=99.3, median=99.8, upper quartile=99.9, upper whisker=100.2}
        ] coordinates {}; 
        \addplot+[
boxplot prepared={lower whisker=98.4, lower quartile=99.3, median=99.7, upper quartile=99.9, upper whisker=100.4}
        ] coordinates {}; 
        \addplot+[
boxplot prepared={lower whisker=90.9, lower quartile=96.8, median=99.7, upper quartile=100.8, upper whisker=106.7}
        ] coordinates {}; 
        \addplot+[
boxplot prepared={lower whisker=77, lower quartile=93.6, median=99.9, upper quartile=104.7, upper whisker=121.1}
        ] coordinates {}; 
        \addplot+[
boxplot prepared={lower whisker=89.7, lower quartile=97, median=99.9, upper quartile=101.925, upper whisker=109.3}
        ] coordinates {}; 
        \addplot+[
boxplot prepared={lower whisker=90.4, lower quartile=97.5, median=99.9, upper quartile=102.3, upper whisker=109.4}
        ] coordinates {}; 
        \addplot+[
boxplot prepared={lower whisker=90.4, lower quartile=97.1, median=99.9, upper quartile=101.6, upper whisker=108.3}
        ] coordinates {}; 
        \addplot+[
boxplot prepared={lower whisker=0, lower quartile=54.5, median=98.9, upper quartile=127.6, upper whisker=237.2}
        ] coordinates {}; 
        \addplot+[
boxplot prepared={lower whisker=0, lower quartile=51.1, median=99.1, upper quartile=130.8, upper whisker=250.3}
        ] coordinates {}; 
        \addplot+[
boxplot prepared={lower whisker=0, lower quartile=42.9, median=99.1, upper quartile=131.9, upper whisker=265.3}
        ] coordinates {}; 
        \addplot+[
boxplot prepared={lower whisker=0, lower quartile=49.5, median=99.4, upper quartile=134.8, upper whisker=262.7}
        ] coordinates {}; 
        \addplot+[
boxplot prepared={lower whisker=0, lower quartile=44.1, median=99.2, upper quartile=134.8, upper whisker=270.8}
        ] coordinates {};


    \end{axis}
\end{tikzpicture}

%% file: images/memory/cpu_quantile_sim.tex
\pgfplotsset{compat=1.12}

\begin{tikzpicture}[font=\sffamily, fill opacity = 0.75, opacity = 0.75]

\pgfplotsset{
boxplot/every box/.style={valuecolour, fill=valuecolour, semithick},
boxplot/every median/.style={valuecolour!90!black,  thick},
boxplot/every whisker/.style={valuecolour, thick},
xtick={1,2,3,4,5,6,7,8,9},
scale only axis=true,
width=0.78\textwidth,
height=0.52\textwidth,
}

    \begin{axis}[colormap/viridis,
        ybar,
        xlabel={Time $t'$ [h]},
        ylabel={CPU [\si{\percent}]}, 
        legend pos=north west,
        ymajorgrids=true, 
        xmajorgrids=true,
        grid style=dashed,
        bar width=0.075cm,
        ymin=1,
        ymax=400,
        xmin=0.5,
        xmax=9.5,
        legend cell align={left},
        boxplot,
        boxplot/draw direction = y,
        table/y=m,
        ]
        
        \addplot+[
boxplot prepared={lower whisker=59.6, lower quartile=79.675, median=98.5, upper quartile=100, upper whisker=102.6}
        ] coordinates {}; 
        \addplot+[
boxplot prepared={lower whisker=0, lower quartile=0, median=86.5, upper quartile=182.4, upper whisker=331.4}
        ] coordinates {}; 
        \addplot+[
boxplot prepared={lower whisker=0, lower quartile=0, median=80.3, upper quartile=194.025, upper whisker=338.1}
        ] coordinates {}; 
        \addplot+[
boxplot prepared={lower whisker=0, lower quartile=0, median=84, upper quartile=186.525, upper whisker=342.6}
        ] coordinates {}; 
        \addplot+[
boxplot prepared={lower whisker=0, lower quartile=0, median=82, upper quartile=226.4, upper whisker=338.6}
        ] coordinates {}; 
        \addplot+[
boxplot prepared={lower whisker=0, lower quartile=0, median=89.9, upper quartile=230.6, upper whisker=350.3}
        ] coordinates {}; 
        \addplot+[
boxplot prepared={lower whisker=0, lower quartile=0, median=74.65, upper quartile=235.6, upper whisker=347.9}
        ] coordinates {}; 
        \addplot+[
boxplot prepared={lower whisker=0, lower quartile=0, median=84.75, upper quartile=231.9, upper whisker=338.5}
        ] coordinates {}; 
        \addplot+[
boxplot prepared={lower whisker=0, lower quartile=0, median=86.8, upper quartile=231, upper whisker=329.9}
        ] coordinates {};


    \end{axis}
\end{tikzpicture}

%% file: images/memory/mem_quantile_poly.tex
\pgfplotsset{compat=1.12}

\begin{tikzpicture}[font=\sffamily, fill opacity = 0.75, opacity = 0.75]

\pgfplotsset{
boxplot/every box/.style={valuecolour, fill=valuecolour, semithick},
boxplot/every median/.style={valuecolour!90!black,  thick},
boxplot/every whisker/.style={valuecolour, thick},
scale only axis=true,
width=0.78\textwidth,
height=0.52\textwidth,
xtick={1,2,3,4,5,6,7,8,9,10,11,12,13,14}
}

    \begin{axis}[colormap/viridis,
        ybar,
        xlabel={Time $t'$ [h]},
        ylabel={Memory [\si{\mega\byte}]}, 
        legend pos=north west,
        ymajorgrids=true, 
        xmajorgrids=true,
        grid style=dashed,
        bar width=0.075cm,
        ymin=1,
        ymax=60,
        xmin=0.5,
        xmax=14.5,
        legend cell align={left},
        boxplot,
        boxplot/draw direction = y,
        table/y=m,
        ]
        
        \addplot+[boxplot prepared={lower whisker=11.117, lower quartile=13.5105, median=15.445, upper quartile=15.482, upper whisker=15.703}
        ] coordinates {}; 
        \addplot+[boxplot prepared={lower whisker=15.758, lower quartile=15.688, median=15.688, upper quartile=15.688, upper whisker=15.676}
        ] coordinates {}; 
        \addplot+[boxplot prepared={lower whisker=15.359, lower quartile=15.359, median=15.359, upper quartile=15.363, upper whisker=15.363}
        ] coordinates {}; 
        \addplot+[boxplot prepared={lower whisker=15.609, lower quartile=15.609, median=15.609, upper quartile=15.613, upper whisker=15.613}
        ] coordinates {}; 
        \addplot+[boxplot prepared={lower whisker=15.367, lower quartile=15.645, median=15.68, upper quartile=15.918, upper whisker=15.957}
        ] coordinates {}; 
        \addplot+[boxplot prepared={lower whisker=15.277, lower quartile=15.684, median=15.863, upper quartile=15.98, upper whisker=16.137}
        ] coordinates {}; 
        \addplot+[boxplot prepared={lower whisker=15.215, lower quartile=15.594, median=15.902, upper quartile=16.016, upper whisker=16.246}
        ] coordinates {}; 
        \addplot+[boxplot prepared={lower whisker=15.605, lower quartile=15.926, median=16.172, upper quartile=16.309, upper whisker=16.418}
        ] coordinates {}; 
        \addplot+[boxplot prepared={lower whisker=15.379, lower quartile=15.766, median=16.082, upper quartile=16.203, upper whisker=16.426}
        ] coordinates {}; 
        \addplot+[boxplot prepared={lower whisker=0.199, lower quartile=15.605, median=21.141, upper quartile=30.133, upper whisker=34.336}
        ] coordinates {}; 
        \addplot+[boxplot prepared={lower whisker=0.594, lower quartile=15.594, median=21.289, upper quartile=30.418, upper whisker=34.656}
        ] coordinates {}; 
        \addplot+[boxplot prepared={lower whisker=1.113, lower quartile=18.062, median=29.371, upper quartile=38.648, upper whisker=52.504}
        ] coordinates {}; 
        \addplot+[boxplot prepared={lower whisker=0.625, lower quartile=18.426, median=29.676, upper quartile=39.129, upper whisker=52.855}
        ] coordinates {}; 
        \addplot+[boxplot prepared={lower whisker=0.062, lower quartile=18.102, median=29.684, upper quartile=38.625, upper whisker=52.32}
        ] coordinates {};


    \end{axis}
\end{tikzpicture}

%% file: images/memory/mem_quantile_sim.tex
\pgfplotsset{compat=1.12}

\begin{tikzpicture}[font=\sffamily, fill opacity = 0.75, opacity = 0.75]

\pgfplotsset{
boxplot/every box/.style={valuecolour, fill=valuecolour, semithick},
boxplot/every median/.style={valuecolour!90!black,  thick},
boxplot/every whisker/.style={valuecolour, thick},
scale only axis=true,
width=0.78\textwidth,
height=0.52\textwidth,
xtick={1,2,3,4,5,6,7,8,9}
}

    \begin{axis}[colormap/viridis,
        ybar,
        xlabel={Time $t'$ [h]},
        ylabel={Memory [\si{\mega\byte}]}, 
        legend pos=north west,
        ymajorgrids=true, 
        xmajorgrids=true,
        grid style=dashed,
        bar width=0.075cm,
        ymin=1,
        ymax=20,
        xmin=0.5,
        xmax=9.5,
        legend cell align={left},
        boxplot,
        boxplot/draw direction = y,
        table/y=m,
        ]

        \addplot+[
boxplot prepared={lower whisker=10.738, lower quartile=13.1992, median=14.809, upper quartile=14.931, upper whisker=15.152}
        ] coordinates {}; 
        \addplot+[boxplot prepared={lower whisker=15.828,  lower quartile=15.828, median=15.828, 
        upper quartile=15.828, 
        upper whisker=15.824}
        ] coordinates {};
        \addplot+[
boxplot prepared={lower whisker=15.613, lower quartile=15.59, median=15.59, upper quartile=15.59, upper whisker=15.359}
        ] coordinates {}; 
        \addplot+[
boxplot prepared={lower whisker=15.656, lower quartile=15.645, median=15.645, upper quartile=15.645, upper whisker=15.312}
        ] coordinates {}; 
        \addplot+[
boxplot prepared={lower whisker=16.281, lower quartile=16.277, median=16.277, upper quartile=16.277, upper whisker=16.273}
        ] coordinates {}; 
        \addplot+[
boxplot prepared={lower whisker=15.926, lower quartile=15.922, median=15.922, upper quartile=15.922, upper whisker=15.918}
        ] coordinates {}; 
        \addplot+[
boxplot prepared={lower whisker=15.438, lower quartile=15.734, median=15.934, upper quartile=15.934, upper whisker=15.988}
        ] coordinates {}; 
        \addplot+[
boxplot prepared={lower whisker=15.953, lower quartile=15.953, median=16.148, upper quartile=16.148, upper whisker=16.344}
        ] coordinates {}; 
        \addplot+[
boxplot prepared={lower whisker=15.949, lower quartile=16.094, median=16.152, upper quartile=16.207, upper whisker=16.27}
        ] coordinates {};


    \end{axis}
\end{tikzpicture}

%% file: sections/conclusion.tex
\section{Conclusion}\label{ss:Conc} 

We proposed and implemented a general algorithm  for  building a Parametric Location Tree for HPnGs with an arbitrary but finite number of general transition firings and presented the computation of transient probabilities in three stages. First the candidate locations, i.e., parametric locations the system can be in at time $t'$, were obtained. Second, the potential domain of all candidate locations was restricted, such that only those values of the random variables remained for which the system certainly  can be in that location at time $t'$. Third, the probability to be in a specific candidate location at time $t'$ was computed by integrating over the joint probability density function, and multiplying the result with the accumulated conflict probability.  We presented two approaches to compute the areas of integration, either using interval arithmetic or using the geometric representation of regions. Both methods then  allow direct integration using Monte-Carlo simulation on intervals or convex polytopes, respectively. Alternatively, the resulting objects can be transformed into standard simplices, which then allows the use of other multi-dimensional integration methods, which is however out of the scope of this paper. 

The feasability study shows that the interval approach performs better for higher dimensions than the geometrical approach. This is due to the dedicated implementation of Fourier-Motzkin which is tailored towards the specific requirements of our algorithms. In contrast the library \textsc{HyPro} is more general and can be used to model check more complex properties. Both geometrical integration methods however require the transformation of the halfspace-representation into the vertex-representation of polytopes which can be exponentional in the state space dimension. 


A case study on a battery-backup system showed the feasibility and current limitations of the approach.

\add{After a validation of the computed probabilities, we performed dedicated measurements for a range of parameter settings. It could be seen that 
the presented geometric integration approaches are not able to perform in as many dimensions as the presented interval integration. Since the geometric approaches can be used to compute not only the transient probability, but also more complex properties, like e.g. model checking. Hence, we performed further measurements on CPU- and memory-usage for the geometric integrations.}

\add{These measurements have shown that this bottleneck currently lies at the pre-computations required for integration, i.e.,  removing time from the state representation, computing bounding boxes, and the inherent changes between $\mathcal{H}$- and $\mathcal{V}$-representations and vice versa. }

We plan to further improve the efficiency of the computation, \add{by better integrating the use of third party libraries.} \remove{e.g., by} \add{Furthermore, we will investigate} a property-guided exploration of the state space and by investigating how multi-dimensional integration can be performed without the need to compute the convex hull.